\documentclass[12pt,english]{article}
\pdfoutput=1

\usepackage[authordate,
giveninits,
backend=biber,
doi=false,
isbn=false,
sorting=nyt,
maxcitenames=2,
minbibnames=7,
maxbibnames=7,
sortcites=true]{biblatex-chicago}

\bibliography{bib/references.bib}

\AtEveryBibitem{\clearlist{note}\clearlist{language}\clearlist{issn}} 
\AtEveryBibitem{%
	\ifentrytype{online}{%
		\clearfield{urlyear}
		\clearfield{urlmonth}
		\clearfield{urlday}
		\clearfield{note}
		\clearlist{language}
	}{%
		\clearfield{eprint}%
		\clearfield{urlyear}
		\clearfield{urlmonth}
		\clearfield{urlday}
		\clearfield{note}
		\clearlist{language}
	}
}%

\usepackage{mathptmx}
\usepackage{microtype}
\usepackage{amsmath}
\usepackage{bbm}
\usepackage[utf8]{inputenc}
\usepackage[T1]{fontenc}
\usepackage{babel}
\usepackage{setspace}
\usepackage{graphicx}
\usepackage{booktabs,dcolumn}
\usepackage{array}
\usepackage{multirow}
\usepackage[usenames,dvipsnames,svgnames,table]{xcolor}
\usepackage[export]{adjustbox}[2011/08/13]
\usepackage{enumitem}
\usepackage[text={16cm,24cm}]{geometry}
\usepackage{ragged2e}
\usepackage{csquotes}

\usepackage[hang, flushmargin, bottom, symbol]{footmisc}
\usepackage[colorlinks=true, linkcolor=blue, citecolor=blue, plainpages=false, pdfpagelabels=true, urlcolor=blue]{hyperref}
\usepackage{float}

\usepackage{subcaption}
\usepackage{caption}
\makeatletter

\newcolumntype{L}[1]{>{\raggedright\let\newline\\\arraybackslash\hspace{0pt}}m{#1}}
\newcolumntype{C}[1]{>{\centering\let\newline\\\arraybackslash\hspace{0pt}}m{#1}}
\newcolumntype{R}[1]{>{\raggedleft\let\newline\\\arraybackslash\hspace{0pt}}m{#1}}

\newcommand{\sym}[1]{\ifmmode^{#1}\else\(^{#1}\)\fi}

\usepackage{calc}
\usepackage{footnotebackref}
\newcommand{\PAPERKEYWORDS}{\textbf{Keywords}: Early Childhood, Height, Reference Points, Nutrition, Anthropometrics}
\newcommand{\PAPERJEL}{\textbf{JEL}: I15, D8, D9, O15}

\newcommand{\PAPERTITLE}{You are What Your Parents Expect:\\Height and Local Reference Points}

\newcommand{\AUTHORWANG}{Fan Wang}
\newcommand{\AUTHORWANGURL}{https://orcid.org/0000-0003-2640-5420}
\newcommand{\AUTHORWANGINFO}{\href{\AUTHORWANGURL}{\AUTHORWANG}: Department of Economics, University of Houston, Houston, Texas, USA (email: fwang26@uh.edu)}

\newcommand{\AUTHORPUENTES}{Esteban Puentes}
\newcommand{\AUTHORPUENTESURL}{https://orcid.org/0000-0001-7603-2197}
\newcommand{\AUTHORPUENTESINFO}{\href{\AUTHORPUENTESURL}{\AUTHORPUENTES}: Department of Economics, Universidad de Chile, Santiago, Chile (email: epuentes@fen.uchile.cl)}

\newcommand{\AUTHORCUNHA}{Flávio Cunha}
\newcommand{\AUTHORCUNHAURL}{https://orcid.org/0000-0003-3265-7894}
\newcommand{\AUTHORCUNHAINFO}{\href{\AUTHORCUNHAURL}{\AUTHORCUNHA}: Department of Economics, Rice University, Houston, Texas, USA (email: Flavio.Cunha@rice.edu)}

\newcommand{\AUTHORBEHRMAN}{Jere R. Behrman}
\newcommand{\AUTHORBEHRMANURL}{http://orcid.org/0000-0001-7835-6283}
\newcommand{\AUTHORBEHRMANINFO}{\href{\AUTHORBEHRMANURL}{\AUTHORBEHRMAN}: Departments of Economics and Sociology and Population Studies Center, University of Pennsylvania, Philadelphia, Pennsylvania, USA (email: jbehrman@sas.upenn.edu)}

\newcommand{\ACKNOWLEDGMENTS}{
We thank Manuela Angelucci, Orazio Attanasio, Elisabetta Aurino, Gabriella Conti, Andrew Foster, Chao Fu, Douglas Gollin, Sonya Krutikova, Elaine Liu, Erzo Luttmer, Cristian Pop-Eleches, Gautam Rao, Mark Rosenzweig, Oded Stark, Aryeh Stein, Chris Taber, and Robert Topel for helpful comments and suggestions. We gratefully acknowledge support from Grand Challenges Canada. Behrman acknowledges support from the Eunice Kennedy Shriver National Institute of Child Health and Human Development (NICHD R21 HD097576). Cunha acknowledges support from National Institute of Health grant 1R01HD073221-01A1. Puentes acknowledges support from the ANID – Millennium Science Initiative Program – Code: NCS17\_015.}

\newcommand{\PAPERABSTRACT}{
Recent estimates are that about 150 million children under five years of age are stunted, with substantial negative consequences for their schooling, cognitive skills, health, and economic productivity. Therefore, understanding what determines such growth retardation is significant for designing public policies that aim to address this issue. We build a model for nutritional choices and health with reference-dependent preferences. Parents care about the health of their children relative to some reference population. In our empirical model, we use height as the health outcome that parents target. Reference height is an equilibrium object determined by earlier cohorts' parents' nutritional choices in the same village. We explore the exogenous variation in reference height produced by a protein-supplementation experiment in Guatemala to estimate our model’s parameters. We use our model to decompose the impact of the protein intervention on height into price and reference-point effects. We find that the changes in reference points account for 65\% of the height difference between two-year-old children in experimental and control villages in the sixth annual cohort born after the initiation of the intervention.\\
\PAPERJEL}

\newcommand{\PAPERDOIURL}{https://doi.org/10.1016/j.jeconom.2021.09.020}
\newcommand{\PAPERINFO}{
This paper has been accepted for publication at the Journal of Econometrics: \url{\PAPERDOIURL}.
}

\begin{document}

\title{
\vspace*{-0.5em}
\onehalfspacing
\PAPERTITLE
\thanks{
\PAPERINFO
}}

\author{
\AUTHORWANG{},
\AUTHORPUENTES{},
\AUTHORBEHRMAN{},
\AUTHORCUNHA\thanks{
\AUTHORWANGINFO;
\AUTHORPUENTESINFO;
\AUTHORBEHRMANINFO;
\AUTHORCUNHAINFO.
\ACKNOWLEDGMENTS
}}

\date{April 1, 2022}

\maketitle

\vspace*{-1em}
\begin{abstract}
\vspace*{-0.5em}
\singlespacing
\PAPERABSTRACT
\end{abstract}
\vfil
\hfil \small\PAPERKEYWORDS \hfil
\vfil
\thispagestyle{empty}
\clearpage

\pagenumbering{arabic}
\setcounter{page}{1}
\renewcommand*{\thefootnote}{\arabic{footnote}}

\section{Introduction}

Insufficient height and weight growth still affect many children around the globe. Estimates are that about 150 million (22 percent) of children under five years of age are stunted \autocite{fao_state_2019}.\footnote{Stunted children have height-for-age more than two standard deviations below the median for a well-nourished population. In 2018, 59 million (30 percent) and 82 million (23 percent) of African and (non-Japanese) Asian children under five years of age were stunted, respectively \autocite{fao_state_2019}.} Studies suggest that these children are at risk of not developing their full human capital potential \autocite{behrman_nutritional_2009, black_early_2017, hoddinott_effect_2008, hoddinott_adult_2013, hoddinott_economic_2013, maluccio_impact_2009, richter_investing_2017, victora_maternal_2008}. Our paper contributes to a growing literature that investigates how interventions in early childhood can contribute to foster human-capital formation for at-risk children \autocite[e.g.,][]{carneiro_heckman_2003, cunha_heckman_2007, cunha_heckman_jeea, cunha_heckman_schennach_2010, heckman_etal_jpube2010, heckman_etal_qe2010, heckman_etal_aer2010, campbell_etal_2014, gertler_etal_2014}.

In general, anthropometric measures, such as height and weight, are partly determined by genes.\footnote{For individuals of European ancestry, genes explain only 10\% of the variability of adult height \autocite{lango_allen_hundreds_2010, berndt_genome-wide_2013}.} But in many cases, variability in anthropometrics due to race or ethnicity is negligible among children who are raised in favorable environments and born to mothers whose nutritional and health needs are met \autocite{habicht_height_1974}. In contrast, environmental factors are significant determinants of growth in the first two years of life \autocite{martorell_intergenerational_2012}. In this paper, we study the determinants of one such environmental factor: protein intake.

Stunting rates are generally higher in locations in which families feed their young children with staple foods that have low protein density because of their availability or affordability \autocite{dewey_reducing_2016}. In these regions, policies that supplement food via lipid-based nutrient supplementation  or that increase parental resources may improve infant outcomes \autocite [e.g., see][]{dewey_reducing_2016, de_groot_cash_2017}. However, stunting also occurs in locations in which animal-source foods are available and affordable \autocite{penny_effectiveness_2005}. This finding suggests that factors other than family resources or prices of foods rich in protein play an important role in determining malnutrition in general and stunting in particular. In this paper, we study nutritional choices by taking into consideration both parental resources as well as locally-determined parental perceptions of normal growth.

Recent successful policies in the prevention of child stunting include actions to influence parental perceptions of normal growth. \textcite{marini_standing_2017} provide a comprehensive description of how Peru successfully reduced stunting rates by 50 percent between 2007 and 2016. We emphasize two initiatives that may have contributed to shifting perceptions. First, UNICEF and the World Bank disseminated a video that communicated height standards that were easy to understand.\footnote{\href{https://www.youtube.com/watch?v=mJieb2Xgt9U}{Video Link}. Due to the perceived impact, the World Bank produced similar videos for other countries.}

Second, the UNICEF ``Good Start'' Program was extended to the entire country.\footnote{See \textcite{lechtig_decreasing_2009} for an evaluation of the UNICEF's ``Good Start'' Program.} The government trained health professionals in local clinics to assess each child's weight and height monthly and plot the information on growth charts.\footnote{The ``Good Start'' Program divided growth charts into two regions: red (indicating undernutrition) and green (indicating normal nutritional status). This simple visual chart contrasts with other growth charts that use percentile information that many parents do not easily understand \autocite[see evidence in][]{ben-joseph_parents_2009}.}$^{,}$\footnote{UNICEF and various ministries of health have advocated such growth monitoring for decades, but until recently there has been little systematic evidence of much if any effects of such efforts \autocite[e.g., see][]{ruel_growth_1992}} The health professionals used the visual information to inform parents about their children's growth status and prescribe corrective actions if growth had not met targets. While illustrative of the forces studied in our paper, the Peruvian case cannot be used to identify the importance of shifting parental and societal norms about ``normal'' growth because the country simultaneously implemented many different interventions.\footnote{See \textcite{marini_standing_2017} and \href{http://blogs.worldbank.org/health/three-factors-halving-childhood-stunting-peru-over-just-decade?cid=EXT\_WBBlogSocialShare\_D\_EXT}{World Bank (2016)} for a helpful summary of all of the programmatic actions that may have contributed to the reduction in stunting.}

Evidence from Africa confirms the importance of parental perceptions in influencing feeding practices and reducing stunting rates. \textcite{fink_home-and_2017} evaluate an intervention in which villages in rural Zambia were assigned to one of three mutually exclusive groups: (1) a control group; (2) a community-meeting group; and (3) a growth-chart group. In the community meeting, children's height and weight were measured, and parents received information about feeding practices that promoted healthy growth. Parents in the third group had a full-size growth chart installed in their home to track children's growth. The chart had a simple design (red if children were stunted for their ages, green if not)\footnote{See \textcite{marini_standing_2017} for the design used in Peru and \textcite{fink_home-and_2017} for the design used in Zambia.} and contained information about how feeding practices could influence children's healthy growth. The authors show that parents' reports of protein intake increased in both intervention groups. However, stunting rates were reduced by 22 percentage points (from 94\% to 72\%) for children of the parents who received the growth charts, but not for children with parents in the community meetings.

The evidence described above justifies developing a model that incorporates parental perceptions of ``normal'' human capital, whether health or skills. We write a model where parental preferences depend on the parents' reference for child health similar to \textcite{kahneman_prospect_1979, bell_1985, Loomes_Sugden_1986, Koszegi_Rabin_2006}. Our empirically guided parameterization of preferences introduces an asymmetry in responses in the sense that parents concerned that their child may not reach the parents' perceived height milestone by age 24 months will behave differently than parents of otherwise identical children who have lower perceptions about height milestones. This asymmetry is documented in the medical literature with respect to children's health \autocite{may_child-feeding_2007, laraway_parent_2010, mathieu_parental_2010, moore_exploring_2012, swyden_relationship_2015, almoosawi_parental_2016}.

In our model, parents take the expected height at age 24 months as the reference point. However, unlike other models of reference-dependent preferences with an expectation-based referent \autocite[e.g.,][]{bell_1985, Loomes_Sugden_1986, Koszegi_Rabin_2006}, the expected height is a normal random variable from the parents' point of view. We assume that parents of children born in period \(y\) observe the height of the children born in period \(y - 1\) and use this information to form their subjective mean and variance beliefs about the reference height. This assumption is consistent with the results reported in \textcite{hansen_generational_2014} who show that changes in the development of children across cohorts affected parental perceptions of normal development and their reports about the developmental status of their children. Also, it aligns with the evidence from research in economics, medicine, and anthropology that finds that parents observe older siblings, other children in the family, or their friends' children to infer what constitutes ``normal'' height and weight \autocite[see, e.g.,][]{reifsnider_mothers_2000, lucas_systematic_2007, thompson_whatever_2014, liu_relative_2016}.\footnote{Parents also report comparing children's clothing to clothing recommended for their children's ages.} Because parents do not realize that they observe a selected sample of children, they may form biased beliefs.

In our framework, biased mean beliefs and uncertainty (i.e., strictly positive variance beliefs) about reference points cause parents to underinvest in their children's human capital. For example, the scoping review by \textcite{cuomo_etal_2019} reports that lack of parental knowledge about development milestones is one of the determinants of tardy diagnoses and treatment of developmental delays. Similarly, \textcite{mulcahy_uncertainty:_2016} show that uncertainty about whether children are following developmental norms is a significant reason parents wait before they seek early childhood intervention services. However, if there are critical or sensitive periods of development, such underinvestment even if temporary may cause permanent damages to children's human capital formation \autocite{cunha_heckman_2007, victora_maternal_2008, victora_worldwide_2010}. This implication from our model justifies the existence of interventions, such as those in Peru and Zambia we describe above, that aim to improve parental knowledge about child development. These interventions reduce biases in mean beliefs, diminish uncertainty about reference points, and raise the levels of investments.

Households may have reference-dependent preferences, biased mean beliefs, uncertainty, and low family income. Because of the low incomes, governments may subsidize the nutrient prices. In our framework, policies that provide nutritional subsidies have three possible effects: (1) the direct impact of subsidies on treated children; (2) the indirect effect of subsidies via shifting reference points on treated children; and (3) the indirect impact of shifting reference points on untreated children. The second and third effects are ``spillover'' effects that have dynamic implications across cohorts. Thus, research that estimates the direct and spillover effects is significant for the design of policies. 

We use data from The Institute of Central America and Panama (INCAP) nutritional trial to estimate the model described above for height. There were four participant villages; two were randomly selected to receive a high-protein supplement, while the other two received a supplement devoid of proteins. In the data, we observe increasing height and nutritional input gaps between villages that received and did not receive protein-rich nutritional supplements. We take advantage of the exogenous variation in protein intake and reference points generated by the experimental design to estimate our model's structural parameters.

We use the estimated model to conduct a decomposition exercise to quantify the mechanisms of the protein-supplementation experiment. We find that reference-point changes account for up to 65\% of the more than one and a half centimeters in height difference between children in the experimental and the control villages at 24 months of age. We interpret the increasing gaps in height and nutrition with children's ages as substantially coming from changes in reference points in treatment villages.

Our work relies on the premise that parents compare their children to children who grow-up in a similar geographic or socio-economic background. As a result, parents or their children form biased norms that, in turn, cause parents to invest suboptimally. Our work is closest to the research by \textcite{kinsler_parental_2016} who study investments in the human capital of school-age children. Similar to our analysis, \textcite{kinsler_parental_2016} find that parents compare their children's abilities with children from the same school. Because of school segregation, this comparison translates into low investment levels from parents (e.g., hiring tutors) for children at the bottom of the skills distribution. An essential difference between our work and theirs is that we take advantage of a randomized controlled trial that causes exogenous changes in parental norms. We use this exogenous variation to estimate the sensitivity of parental investment behavior to norms.

Our work also relates to the literature on parenting in developmental psychology and economics. The literature in developmental psychology finds that the lower the parents’ socioeconomic status, the lower their expectations about their children's cognitive development \autocite[e.g., see][]{Epstein_1979, hess_et_al1980, ninio1980, niniorinot1988, mansbach_greenbaum_1999}. The literature in economics has shown that parental beliefs about the production function of human capital vary across socioeconomic groups \autocite[][]{cunha_eliciting_2013, boneva_human_2018}. More recent literature uncovers that beliefs predict investments in the human capital of children \autocite[][]{orazio_cunha_jervis_2019, cunha_eliciting_2021}. Finally, scalable parent-directed interventions can influence parental beliefs, which, in turn, impact investments in children \autocite[][]{cgn_2021}. We see our paper as complementary to that literature because our contribution is to incorporate parental beliefs about reference points (or norms) about child developmental outcomes. 

In section 2, we present a model of reference points in a setting of a household deciding inputs for a child outcome, such as health. In Section 3, we describe the data we use to estimate the model. In Section 4, we discuss identification and estimation. In Section 5, we show parameter estimates and present evidence about model fit. Also, we use the estimated model to decompose the impact of the intervention on height. Section 6 concludes.

\section{The Model}

In this section, we develop a model in which parents have to choose a single input for one relevant child outcome. We focus on nutrition as the input and health (height) as the outcome, but the model can be applied to any human-capital input or human-capital outcome developed during childhood and can be extended to multiple inputs and multiple outcomes.

In our model, household choices are functions of prices, incomes, and subjective beliefs about the reference health. Each household solves a static maximization problem after the birth of a child. We assume that households make a single decision about the total nutritional input (i.e., grams of proteins) for their newborns \autocite[e.g., see][]{moradi_nutritional_2010, puentes_early_2016}.\footnote{The model focuses on the development of height that occurs in the first 24 months of a child’s life. The first 1,000 days after conception is a critical period for nutrition and development \autocite{victora_maternal_2008, victora_worldwide_2010}.} However, our framework is general, and we could use the same model to study cognitive or socio-emotional development. In such cases, the input would be the frequency or the quality of the interactions \autocite[e.g., see][]{cunha_heckman_schennach_2010}.

Each household forms subjective beliefs about the reference height by looking at their relevant comparison group. Parents of newborns observe the heights of two-year-old children in their location to estimate subjective beliefs, which they combine with other state variables to choose their children’s nutritional level. A policy that increases height for children in a location will lead parents to update their subjective beliefs, which, in turn, will impact their nutritional decisions. Therefore, such a policy can have dynamic effects because individuals’ choices have aggregate implications that determine the dynamics of subjective beliefs. We describe this idea formally below.

\subsection{Budget}

Let Y denote the household income over the first 24 months of life after birth. Households allocate income to household consumption, $C$, or child nutrition $N$, whose price is $p^{N}_{yv}$:

\begin{equation}
C + p^{N}_{yv}\cdot(1-\delta\cdot\mathbbm{1}\left\{v=Atole\right\})\cdot N = Y \label{eq:budget_constraint}
\end{equation}

We model the protein supplementation policy as a $\delta$ discount on the price of protein in Atole villages, which were randomly assigned to receiving the high-protein supplementation shake.\footnote{As we describe below, the treatment villages received Atole -- high in protein -- and the control villages received Fresco --- the supplement without protein.} This assumption implies that if an Atole village child consumes $N$ grams of protein, the fraction $\delta$ comes from the feeding centers' protein supplements. We argue that this assumption is justifiable because the share of proteins obtained from the feeding centers for each child is positive for all households, is on average 38 percent across cohorts, and is not significantly different across cohorts (F-test p-value 0.63).

\subsection{Production function}

Health (height) at month 24, $H^{24}$ is determined by:

\begin{equation} 
H^{24} = \exp(A+ X\cdot\alpha+\epsilon)\cdot N^{\beta} \label{eq:prod_function}
\thinspace,
\end{equation}

where covariate vector $X$ includes the initial conditions, such as length at birth and gender of the child. The variable $\epsilon$ represents the normally distributed i.i.d. health productivity shock for each child with mean zero and standard deviation $\sigma_{\epsilon}$, which parents observe at the time of choosing $N$. The parameter $A$ relates to the average level of productivity of \(N\) in producing \(H^{24}\), and $\alpha$ represents the impact of covariates on the marginal productivity of $N$. Initial conditions have positive impacts on month-24 height depending on the value of $\alpha$.\footnote{This is a more general specification than a model in which the difference in health between month 0 and month 24 is the production function output and initial health is not included in $X$. Compared to a model with difference in health as the output and that also includes initial health in $X$, the model here produces similar coefficients.} The production function includes protein input $N$ in the first two years, with $\beta$ determining the concavity of the production function with respect to nutritional inputs.

We assume that only total protein intake in the first 24 months matters. Potentially, the timing of nutritional intakes within the first 24 months could be important. In the data, however, lagged inputs over the first 24 months of life are persistent, and it is difficult to distinguish relative productivity across subperiods \autocite{puentes_early_2016}.

In our model, households know the production function. Our assumption contrasts with the literature in which parents have beliefs about the parameters of the production function \autocite[e.g., see][]{cunha_eliciting_2013}. A more general model would allow parents to have biased beliefs about the parameter $\beta$. We do not pursue this approach because the intervention we consider in this paper did not include educating parents on the benefits of protein on health. We return to this issue in Section \ref{sec:estiidentify}.

\subsection{Preferences}

The utility of a household in village $v$ whose child was born in period $y$ -- each period is two years -- is a function of the household consumption, $C$, and the child's height at age 24 months, $H^{24}$:
\begin{equation}\label{eq:preferences}
U_{yv} \left(C, H^{24}, R_{yv} \right) =C+\rho\cdot C^{2}+\gamma \cdot H^{24}+\lambda \cdot (H^{24} - R_{yv}) \cdot \mathbbm{1} \left\{ H^{24} \ge R_{yv} \right\}
\thinspace,
\end{equation}
where $\mathbbm{1}$ denotes the indicator function. 

Parental utility is a function of the reference health, as measured in height, with which households compare their children's health, $R_{yv}$. The parameter \(\gamma\) is positive, but, depending on the relative values of $\lambda$ and $\gamma$, preferences for health are flexible and could be linear, convex, or concave. If $\lambda=0$, preferences are linear in health $H^{24}$. If $\lambda >0$, preferences are convex in health, and parents invest more in health after health exceeds the reference point. If $-\gamma < \lambda<0$, preferences are strictly increasing and concave: parents have strong preference for their child to reach the reference point and weaker preference beyond that. If $-\gamma = \lambda < 0$, households gain utility from increasing health up to the reference health $R_{yv}$, but there are no utility gains from increasing health beyond that point. If $\lambda < -\gamma < 0$, preferences for health peak at the reference point, meaning that households want their children to be healthier (e.g. taller) up to the reference point but not beyond it.

\subsection{Parental Beliefs about Reference Points\label{sec:information}}

In our model, the parents of children born in village $v$ and period $y$ adopt the expected height of children born in the same village in period $y-1$ as their reference point. Thus, $R_{y,v} = \textbf{E}\left(H^{24}_{y-1,v}\right)$. In effect, parents of a new cohort do not know what the average height of the new cohort will be in the future. Instead, they infer it by observing the realized outcomes of the previous cohort.
Therefore, our model is similar to the literature in behavioral economics that expresses reference points as expectations \autocite[][]{bell_1985, Loomes_Sugden_1986, Koszegi_Rabin_2006}. However, we depart from that literature because in our framework parents do not know $\textbf{E}\left(H^{24}_{y-1,v}\right)$. Instead, they have subjective beliefs about it. In particular, our model proposes that, \emph{from the point of view of the parents}, $\textbf{E}\left(H^{24}_{y-1,v}\right)$ is normally distributed with mean $\mu_{R_{y,v}}$ and variance $\sigma_{R_{y,v}}^{2}$. 

We assume that parents use data on the children they observe to estimate $\mu_{R_{y,v}}$ and $\sigma_{R_{y,v}}^{2}$ \autocite{nerlove_adaptive_1958, greenwood_waves_2015}. This assumption about the parental estimation of mean beliefs is plausible given the geographic proximity and the high degree of interactions among households within villages. The assumption that the relevant comparison group corresponds to children from the same village is in line with the literature that has found that the relevant comparison groups are individuals who are close \autocite[e.g., see][]{kinsler_parental_2016,liu_relative_2016}.

Suppose there are $M_{y-1,v}$ children born in period $y-1$ in village $v$. $\left\{H^{24}_{i,y-1,v}\right\}^{M_{y-1,v}}_{i=1}$ are the realized health outcomes observed by parents of newborns in period $y$ to determine the subjective mean and variance beliefs about the reference point $R_{y,v} = \textbf{E}\left(H^{24}_{y-1,v}\right)$. Specifically, for mean beliefs:
\begin{equation}
    \mu_{R_{y,v}} = \frac{1}{M_{y-1,v}} \sum_{i=1}^{M_{y-1,v}} H_{i,y-1,v}^{24} \label{eq:mean_belief}
    \thinspace\thinspace.
\end{equation}
$\mu_{R_{y,v}}$ is a sample mean that differs depending on the realized outcomes of children born in the previous cohort in the village setting. When reference points uncertainty arises from sampling error, variance beliefs would be:
\begin{equation}
    \label{eq:variance_belief}
    \sigma_{R_{y,v}}^{2} = \frac{1}{M_{y-1,v}\left(M_{y-1,v}-1\right)} \sum_{i=1}^{M_{y-1,v}}\left(H_{i,y-1,v}^{24} - \mu_{R_{y,v}}\right)^2 
    \thinspace\thinspace.
\end{equation}

We do not include heterogeneity in $\mu_{R_{y,v}}$ or $\sigma_{R_{y,v}}^{2}$ by socio-economic characteristics because the experimental study took place in very small villages. We expect every household to observe the same height distribution.

In our framework, if a significant fraction of the selected sample has developmental delays, the parents will form biased mean beliefs about reference points, which, in turn, causes sub-optimal investments.\footnote{In our context, we say that beliefs are biased if they deviate from WHO anthropometric standards, which are based on growth data from ``healthy breastfed infants and young children from widely diverse ethnic backgrounds and cultural settings'' who live ``under conditions likely to favor the achievement of their full genetic growth potential''
\autocite{world_health_organization_who_2006}.} This aspect of the model calls for interventions that address such forms of biases. One type of intervention is to improve parental knowledge about developmental norms described in \textcite{marini_standing_2017, fink_home-and_2017}. In addition, as we demonstrate in our decomposition exercise, policies that impact developmental outcomes in the selected group of children will impact the bias in mean beliefs and, consequently, affect investment choices.

\subsection{Maximization Problem\label{sec:maximize}}

Following our discussions on parental beliefs, given $\Omega_{y,v} = \big(Y, p^{N}_{yv}, \delta, X, \epsilon, \mu_{R_{y,v}}, \sigma_{R_{y,v}}^{2} \big)$, each household solves the following maximization problem:

\begin{equation}
\label{eq:optimain}
\max_{C,N} \Bigg\{C+\rho\cdot C^{2}+  \left\{ \gamma\cdot H^{24}+\lambda \cdot \int_{R_{yv}} \left(H^{24}-R_{yv}\right)\mathbbm{1}\left\{ H^{24} > R_{yv}\right\}\phi \left(R_{yv};\mu_{R_{yv}}, \sigma_{R_{yv}}^{2} \right) dR_{yv} \right\} \Bigg\}
\thinspace,
\end{equation}
subject to the budget constraint \eqref{eq:budget_constraint}, the production function \eqref{eq:prod_function}, the equation that determines mean beliefs \eqref{eq:mean_belief}, and some measure of variance beliefs $\sigma^2_{R_{y,v}}$. Let $N\left(\Omega_{y,v}\right)$ denote the policy function for nutrition. If $\sigma^2_{R_{y,v}}=0$, preferences for health would be piecewise linear with a kink at $\mu_{R_{yv}}$. If $\sigma^2_{R_{y,v}}>0$, preferences for health are continuously differentiable; if additionally $\gamma>0$ and $\lambda<0$, preferences for health are concave. 

The optimization problem in Equation \eqref{eq:optimain} does not permit analytical solutions.\footnote{In Appendix Section \ref{sec:apprefsd}, we discuss the first-order condition for the optimal nutritional choice problem. In Appendix Section \ref{sec:modelsolu}, we describe the procedures for numerical solutions.} To illustrate how changes in key parameters impact optimal choices, we present in Figure \ref{fig:indiff} the consumption-health possibility frontier along with several indifference curves for an individual, using estimated parameters from our model.\footnote{The consumption-health possibility frontier is jointly determined by the household budget and the child height production function.} In each panel, solid dots indicate optimal choices (i.e., the combination of height and household consumption where indifference curves are tangent to the consumption-health possibility frontier).

Panel a shows several indifference curves and the consumption-health possibility frontier. The figure clearly shows the asymmetry in indifference curves. To the left of the mean reference height, parents are willing to sacrifice a large amount of consumption for a small increase in the child's height. In contrast, to the right of the mean reference height, parents are willing to forego only a small amount of consumption for a large increment in height. 

In panel b, we vary the $\lambda$ parameter. Marginal benefits of additional nutritional intakes on expected utility are increasing in $\lambda$. Additionally, reference points matter more when $\lambda$ deviates more from zero. Visually, at more negative values of $\lambda$, the curvature of the indifference curve increases. The greater the curvature of the indifference curve, the more critical the role of reference height in determining parental nutritional choices and, consequently, the child's height at age 24 months. 

In panel c, we fix $\lambda$ and vary $\mu_{R_{y,v}}$. When $\lambda<0$, the marginal benefits of additional nutritional intakes are increasing in $\mu_{R_{y,v}}$. The greater the value of $\mu_{R_{y,v}}$, the lower the household consumption and the taller the child at age 24 months. 

Finally, panel d of Figure \ref{fig:indiff} presents the effect of increasing uncertainty about the parental estimates of mean height, $\sigma^2_{R_{y,v}}$. Given the estimated parameters from the empirical model, higher values of $\sigma^2_{R_{y,v}}$ increase the marginal benefits of additional nutrition when expected height exceeds $\mu_{R_{y,v}}$ and reduces the curvature of the indifference curves. These lead to an increase in nutritional choices. In the limit, as $\sigma^2_{R_{y,v}}$ approaches infinity, preferences become linear in height, and predictions from models with and without reference points become empirically indistinguishable.

We use the estimated parameters to produce Figure \ref{fig:indiff}. However, in general, the household response to an increase in uncertainty depends on the values of $\gamma$ and $\lambda$. When marginal utility gains from additional investment in health are positive after the mean reference point $\mu_{R}$ -- which means $-\gamma < \lambda < 0$ -- an increase in $\sigma_{R}$ will lead to an increase in investments in health. In contrast, if marginal utility from additional investment in health is negative after the mean reference point $\mu_{R}$ -- which generates backward bending indifference curves -- a similar increase of $\sigma_{R}$ could reduce investment in health. Ceteris paribus, there is a threshold level of $\lambda$ where increases in $\sigma_{R}$ have no impact on health investments. In Appendix Section \ref{sec:sigmaopti}, we discuss the intuition behind these results and provide graphical illustrations.

\subsection{Evolution of Distribution Function of Height at Age 24 Months\label{sec:evolution}}

As discussed above, static individual choices have dynamic aggregate effects because the heights for the cohort born in period $y$ are realized in period $y+1$ and determine the reference point beliefs for the cohort born in $y+1$. We make this argument formal by presenting how the distribution function of heights at age 24 months evolves. 
Let $\Gamma_{y,v} \left( h^{24} \right)$ denote the distribution function of heights at age 24 months for cohort $y$ in village $v$. Note that the parameters $\mu_{R_{y,v}}$ and $\sigma_{R_{y,v}}$, elements of the state vector $\Omega_{y,v}$, are functions of the distribution function $\Gamma_{y-1,v}$ based on realized heights of the cohort born in period $y-1$. Therefore, we now use the notation $\Omega_{y,v} \left( \Gamma_{y-1,v} \right)$ to make this dependence explicit. The following equation describes the equilibrium law of motion for the distribution function of heights at age 24 months:

\begin{equation}
\label{eq:hmeasure}
    \Gamma_{y,v} \left( h^{24} \right) = \text{Pr} \left[H^{24}\left(N\left(\Omega_{y,v} \left( \Gamma_{y-1,v} \right)\right),X,\epsilon \right) \leq h^{24} \right]
    \thinspace\thinspace,
\end{equation}
where $\Gamma_{y,v}$ is the distribution function of $H^{24}_{y,v}$ in village $v$ for the cohort born in period $y$, and is realized in period $y+1$. Given all other parameters, Equation \eqref{eq:hmeasure} provides the evolution of reference point beliefs and implicitly links the probability distribution functions of $H^{24}_{y,v}$ and $H^{24}_{y-1,v}$. The knowledge of this distribution function is useful because it determines the reference-point parameters. We can use this equation to simulate paths of height distribution over time across villages. We can also explore the same equation to simulate paths for counterfactual policies, which we will need to do to conduct our decomposition exercise. 

\section{Data}

This section describes the data we use to estimate the model presented in the previous section. We present below specific data features that -- in the context of the model -- we can use to isolate the effects of changes in the parameters of the reference points.

\subsection{Survey Design and Sample}

The data we use in this paper comes from an experimental intervention conducted by The Institute of Nutrition of Central America and Panama (INCAP), which started a nutritional-supplementation trial in 1969. Four villages from eastern Guatemala were selected, one pair of villages that was relatively populous (\textasciitilde900 residents each) and one pair that was less populous (\textasciitilde500 residents each). The villages were similar in child nutritional status, measured as the height at three years of age \autocite{habicht_nutritional_1995}. Over 50\% of children lacked proper nutrition and were severely stunted, measured as height-for-age z-scores less than -3.\footnote{Guatemalan children continue to suffer from severe malnutrition. In 2015, among Guatemalan households in the lowest quintile of wealth, approximately 70 percent of children younger than five were stunted. In middle-quintile Guatemalan households, 45 percent of children younger than five were stunted \autocite{fao_state_2019}.} The intervention consisted of randomly assigning nutritional supplements. One large and one small village were selected to receive a high-protein drink called Atole, and the other two were selected to receive an alternative supplement called Fresco. Each serving of Atole (180 ml) contained 11.5 grams of protein and 163 kcal. Fresco had no proteins, and each serving (180 ml) had 59 kcal. The central hypothesis was that the protein supplementation would accelerate physical and mental development \autocite{habicht_nutritional_1995}. The intervention started in February 1969 in the larger villages and in May 1969 in the smaller villages and lasted until the end of February 1977, with data collection taking place until September 1977 \autocite{maluccio_impact_2009,islam_evidence_2009}. The nutritional supplements were distributed in feeding centers located centrally in each village. The centers were open twice a day, two to three hours in the mid-morning and two to three hours in the mid-afternoon. All village members had access to the supplements at the feeding centers.

Information on supplement intake was collected daily for all children up to seven years old. Interviewers collected data on height, home diets, and supplement diets every three months for children between 0 and 24 months. All children reported positive supplement intakes. The home dietary data corresponds to 24-hour recall in the large villages and 72-hour recall in the small villages. It is possible to calculate protein intakes from the home dietary data, which we use in our estimation. Anthropometric measures were collected every three months for children 0 to 24 months old.

Given the quarterly data collection for the INCAP dataset in the first 24 months of life, a child was observed up to nine times. There were 1155 individuals for whom we have at least one height observation between months 0 and 24, and 363 individuals for whom heights were observed nine times. We focus on 503 individuals for whom we have heights at birth, heights at month 24, and at least two observations of nutritional inputs between months 15 and 24.\footnote{For 378 individuals, we observe nutritional intakes in months 15, 18, 21 and 24, for 100 individuals, we observe nutritional intakes three times, and for 25 individuals, we observe nutritional intakes two times.} For these 503 individuals, we have information on household income for one year in the period of analysis. We also have from the INCAP survey food price data measured at wholesalers' purchasing cost per 10,000 grams of each type of food. The food prices are common for Atole and Fresco villages. Using these data, we estimate protein prices from a simple hedonic pricing equation system in which the price for each unit of a food item is determined by the sum of the protein and non-protein caloric values for each unit of food item multiplied by the year-specific protein and non-protein-calorie prices up to a random error term. More information is available upon request.

\subsection{Descriptive statistics \label{sec:stat}}

Table \ref{summcovarmain} presents summary statistics for the variables that we use in our analysis. In Panels a and b, we show statistics on gender, income, and prices for our main sample of 503 individuals (panel a) and gender and income for the fuller sample of 1155 individuals (panel b). In Panels c and d, we show statistics on heights and nutritional intakes, respectively. Table \ref{summcovarmain} has five columns. The first column presents the overall means and standard deviations in Atole and Fresco villages combined. The second and third columns present Atole and Fresco village-specific means and standard deviations. Column four shows the gaps in means between Atole and Fresco villages for each variable, and column five presents the p-values for the statistical significance of these gaps.

As mentioned before, the intervention took place in four villages, two Atole or treatment villages and two Fresco or control villages. In the rest of the paper, when we refer to Atole and Fresco villages, we merge the information of the two villages that received the same supplement. The limited number of villages might impact the descriptive statistics' standard errors since villages can share common unobserved shocks. We follow the methods developed by \textcite{donald2007inference} and \textcite{cameron2015practitioner} to study how robust our results are to this clustering. The method proposed by \textcite{donald2007inference} consists of estimating averages by clusters, controlling for individual variables, and using those averages in the regressions. This method greatly reduces the number of observations. We define clusters as year-village pairs and half-year-village pairs to implement this procedure. Following \textcite{cameron2015practitioner}, we also implement a pair-cluster bootstrap, using the same cluster definition as in the \textcite{donald2007inference} method. Table \ref{summcovarmain} and Figure \ref{fig:PortHgtGap} report the results without using the clusters corrections, but the results are robust to those methods.

Panels a and b show that the survey is well-balanced for gender and income between Atole and Fresco villages. Panel a shows that male children account for 52 percent of our main sample in both Atole and Fresco villages. In panel b, for the larger sample that includes individuals for whom we observe height once in the first two years of their lives, the male share is 53 percent in both Atole and Fresco villages. These indicate that gender-compositional differences do not drive differences between Atole and Fresco villages in height outcomes and nutritional intakes.

Panels a and b also show that annual household incomes for Atole and Fresco villages have comparable distribution functions. \footnote{The survey contains a wealth index constructed with data collected in 1967 and 1975 for all individuals. We also know the mean and the standard deviation of income for 1974 but only at the village level. Assuming that annual household income follows the same distribution as the household wealth distribution and assuming log-normality of the income distribution, we impute annual household income. We multiply annual income by two to calculate each household’s total resources in the first two years of a child’s life.}  In panel a, for our main sample, average annual household incomes are 503 quetzales\footnote{Real terms for 1975; the exchange rate was 1 quetzal for 1 US dollar.} in Fresco villages, and 526 quetzales in Atole villages. The difference is statistically insignificant, with the p-value equal to 0.59. Standard deviations for both villages are almost identical at \textasciitilde460 quetzales, indicating a high similarity in the distribution of incomes between Atole and Fresco villages. For the larger sample in panel b, average incomes per year are almost identical at 454 quetzales in Atole villages and 444 quetzales in Fresco villages, a statistically insignificant difference of only 2 percent. The higher income in panel a’s main sample compared to the income in panel b’s larger sample indicates potential selection in terms of which children were observed more often and consistently. However, both gender and income in both panels a and b are almost precisely balanced, indicating a high similarity between Atole and Fresco villages.

Panel a also presents summary statistics for the village averages of individual average protein prices between months 15 and 24 of age for each child. We calculate annual protein prices based on the average of annual food prices for rice, eggs, chicken, corn, and beef weighted by their respective protein shares \footnote{We obtain protein shares from USDA Food Composition Database \autocite{united_states_department_of_agriculture_usda_2016}, which provides the protein values per 100 grams of various food items.}. By construction, food prices differ for each calendar year but are identical for Atole and Fresco villages. For each child, we average over prices that the child faces in months 15, 18, 21, and 24 of age --- the months over which we calculate average nutritional intakes shown in panel d. Depending on the month and year of birth,\footnote{The birth date distributions in Atole and Fresco villages are shown in panels a and b of Figure \ref{fig:PortHgtGap} where the size of scatter plots indicate the relative sample sizes of birth in each calendar-year between 1970 and 1975.} the average price for each child differs. In Column one of panel a, we show the overall averages of these individual averages, which is 52.58 quetzales per 10k grams of protein. The standard deviation is 3.87 quetzales or 7.4 percent of the means, indicating significant price variations across individuals. Atole and Fresco villages' averages are almost identical at 52.47 and 52.68 quetzales (p-value 0.54 for difference), indicating that the distributions of birthdates between Atole and Fresco villages are well-balanced.

Aggregating across cohorts, panel c of Table \ref{summcovarmain} shows at birth, Atole-village children, with an average height of 49.52~cm, are 0.27~cm shorter on average than Fresco-village children whose average height is 49.79~cm. This difference is not statistically different (p-value 0.19). Moving from birth to month 24, heights for children in Atole villages increase faster than heights for children from Fresco villages. At month 6, Atole children are on average 0.44~cm taller than Fresco children. This gap widens to 0.68~cm at 12 months and 0.92~cm at 18 months of age. At month 24, the average height in Atole villages is 78.30~cm, and the average height in Fresco villages is 76.97~cm --- the Atole height premium is 1.33~cm (p-value \(\le\) 0.005).

Panel d of Table \ref{summcovarmain} presents averages for nutritional intakes per day for children across birth cohorts between 1970 and 1975. We focus on nutritional intakes in the second year of life in months 15, 18, 21, and 24.\footnote{In the first year of life, a significant portion of children obtain nutrition from breast milk, and it is not easy to impute the protein and caloric values of breastmilk given the heterogeneity in breastmilk and feeding durations. We have nutritional intake data for the 503 individuals of the main sample in most of the four feasible survey months.} For month 15 of age, Atole children average 20.07 grams of protein intake per day, 5.78 grams more than children in Fresco villages. In months 18, 21, and 24 of age, the average daily intake gap between Atole and Fresco villages widens to 6.14, 7.82, and 8.56 grams per day. Overall, averaging across the four quarters in the second year of life for each child, average protein intakes in Atole villages at 25.84 grams are 7.06 grams (38 percent) higher than the average for Fresco villages (18.78 grams per day). The final row of panel d shows the village averages of individual average kcal per day of caloric intakes from non-protein sources over months 15, 18, 21, and 24 of age, which is 700.55 kcal per day in Atole villages and 691.78 kcal per day in Fresco villages, a statistically insignificant gap of 2.7 percent.

\subsection{Gaps Across Cohorts \label{sec:gap}}

We now present data on heights at 24 months of age and average protein intakes between months 15 and 24 of age. We display such data across cohorts of children born between 1970 and 1975. As described earlier, the nutritional-supplementation experiments started in the first half of 1969. The 503 children in our sample were born between 1970 and 1975.\footnote{We use birth dates to aggregate children into annual birth cohorts. Out of the 503 children in our sample, 39 were born in 1970, 82 in 1971, 93 in 1972, 96 in 1973, 101 in 1974, and 92 in 1975. The lower number of observations for children born in 1970 is because information on small villages started to be measured on May 1st, 1969, which decreases the number of observations with data at birth for that cohort, compared with the rest.} The families into which the children were born change considerably across cohorts. The percentage of parents in the cohort $t+1$ who were not parents in the $t$ cohort is 83\%, 87\% , 93\%, 92\%, and 95\%, for each year $t$ between 1971 and 1975, respectively.

\subsubsection{Height Gaps Across Cohorts}

Panel a of Figure \ref{fig:PortHgtGap} presents results for cohorts aggregated over each birth year between 1970 and 1975. The average height-at-month-24 gaps between Atole and Fresco children are 0.2~cm ($76.6-76.4$) for the 1970 cohort and 1.6~cm ($78.8-77.1$) for the 1975 birth cohort.\footnote{Children born in the first half of 1970 have approximately the same average height in Atole and Fresco villages at 24 months of age -- both at approximately 76.4~cm. However, for those born in the second half of 1975, the average month-24 heights are 78.9~cm and 76.9~cm for Atole and Fresco village children, respectively.}$^{,}$\footnote{We test whether other variables could explain the increasing heights at month 24 gap across cohorts in Atole and Fresco villages. We regress heights at month 24 on four birth cohort year groups -- 1970, 1971, 1972-73, and 1974-75 -- and the interaction of these birth cohort groups with the Atole dummy. We include gender, protein prices, incomes, and initial heights. These variables are the state variables of our structural model. The results with and without covariates are similar, which is not surprising given that, as we saw in Table \ref{summcovarmain}, there are no significant statistical differences between Atole and Fresco villages in gender ratios, incomes, protein prices, and initial heights.} 

Panel a also shows linear and local polynomial approximated height trends in Atole and Fresco villages. They show a relatively flat pattern for heights at month 24 across cohorts in Fresco villages and a significantly increasing pattern for heights at month 24 across cohorts in Atole villages. Specifically, the linear trend indicates that each additional cohort year is associated with an increase of 0.34~cm (s.e. 0.13) in heights at month 24 in Atole villages and a slightly positive but insignificant increase of 0.11~cm (s.e. 0.14) in heights at month 24 in Fresco villages. 

\subsubsection{Nutritional Gaps Across Cohorts}

Panel b of Figure \ref{fig:PortHgtGap} shows protein intakes across cohorts. Again, we show village cohort averages aggregated over individual averages for 15, 18, 21, and 24 months. We aggregate results to full-year cohorts in panel b, which shows that the average protein intake gaps between Atole and Fresco villages were 3.9 grams ($21.3-17.4$) in 1970 and 7.9 grams ($26.3-18.4$) in 1975.\footnote{Atole children born in the first half of 1970 have 21.1 grams of daily protein on average, and corresponding Fresco children had 17.4 grams on average (3.7 grams gap). For those born in the second half of 1975, the cohort-group average increased to 27.3 grams per day in Atole villages and 18.9 grams in Fresco villages (8.4 grams gap).} Looking at percentage differences, for the annual cohorts, the average protein intakes in Atole villages were 22, 36, 37, 35, 40, and 43 percent higher in Atole villages than in Fresco villages for the six annual birth cohorts from 1970 to 1975.\footnote{We test for the significance of the average protein-intake gap between Atole and Fresco villages across cohorts. We include gender dummies, food prices, incomes, and initial heights as covariates. Without controls, the average intake gaps were 3.85 (1970), 6.76 (1971), 6.70 (1972-73), and 8.02 (1974-75) grams per day between Atole and Fresco village cohorts. Including controls, the gap estimates were 4.31, 7.35, 6.88, and 7.88 grams per day,  showing a similar increasing trend.}

Panel b shows trends from linear and local polynomial approximations of average protein intakes across cohorts similar to those for heights. Specifically, we find that each additional year is associated with a 0.68 grams (s.e. 0.38) increase in intakes for Atole children and an insignificant increase of 0.17 (s.e. 0.26) grams intakes for Fresco children.

Figure \ref{fig:PortHgtGap} shows that for successive cohorts from 1970 to 1975, there were generally increasing protein-intake gaps that correspond to increasing height gaps between Atole and Fresco villages. The overall (i.e., aggregated across all cohorts) Atole and Fresco protein-intake and height gaps have been observed before \autocite[see, e.g.,][]{puentes_early_2016}, but not the increasing protein-input and height gaps across cohorts.

The empirical question that we face is what can explain these increasing gaps between Atole and Fresco villages. We discussed that gender shares, incomes, prices, and initial heights do not differ significantly between Atole and Fresco villages and do not seem to explain the increasing differences between Atole and Fresco villages across cohorts. Local feeding centers carried out no changes in the protein and non-protein supplementation policies that might have impacted cohorts differentially over time. Our structural model with reference-dependent utilities can explain these increasing protein-intake and height gaps between Atole and Fresco villages.

\subsection{Alternative Theories}

Potentially, alternative theories could also be consistent with the observed height and protein gaps. Households might slowly begin to trust the supplement and slowly increase the intakes of Atole and Fresco. Over time, this could lead to growing calories consumption in both villages, which we observe, and the growing protein consumption gap across villages. These facts are consistent with both the reference-point theory and the trust theory. But since the supplement is available daily, we would expect trust to be developed in a matter of weeks or months, not slowly over almost a decade. Additionally, the trust theory should increase the extensive margin, but as discussed earlier, all children have positive supplement intakes since the first year of the experiment.\footnote{Defining trust as the ``subjective probability individuals attribute to the possibility of being cheated'', \textcite{guiso_sapienza_zingales_2008_JoF} find that a lack of trust impact both extensive and intensive margins of consumer (investor) choices.}

Another theory that might be consistent with the gaps we find in the data is that the supplements allow households to learn about the production function. Under a learning model, specific configurations of mean and variance parental beliefs could generate similar dynamics; however, the opposite dynamics might arise if parents overestimate the impact of protein on height or face uncertainty about the impact of protein on height. If parents overestimate the impact of protein on height, the consumption of protein decreases over time as parents converge down to the objective estimates of the impact of protein. If parents face uncertainty about the impact, they will experiment and give more protein to learn and reduce their uncertainty about the impacts of protein on height. In other words, the learning model might not predict the same qualitative path of adjustment of protein consumption and evolution of height over time \autocite[e.g., see][]{doi:10.1086/663356}.

Finally, preferences with habit formation could produce similar dynamics in protein intakes.\footnote{In his classic work, \textcite{pollak_1970} defines habit formation as ``allowing past consumption to influence current tastes''.} Habit formation would apply to both control and treatment villages because the villages only differed in the type of nutritional supplementation they received: Fresco and Atole. The relatively significant increase in average caloric consumption from the supplement in the Atole villages between 1970 and 1975 of 47 kcal per day in comparison with the average increase of 11 kcal per day in the Fresco villages may suggest that something beyond habit formation happened in the Atole villages in contrast to the Fresco villages. However, we cannot rule out habit formation as an alternative explanation. We also note that habit formation and reference points are not mutually exclusive theories, so they could simultaneously operate. 

Despite our firm beliefs that these explanations are unlikely to drive the dynamics we observe in this data, we recognize that these alternative theories cannot be directly rejected with the information we have in our data. A definite test of the different mechanisms implied by these models would require researchers to elicit parental trust about public programs, parental beliefs about reference points, and parental perceptions about height production functions. To the best of our knowledge, no parenting program to date has collected all of this information simultaneously in surveys with participating parents.

\section{Identification} \label{sec:estiidentify}

In this section, we discuss how we explore the experimentally-induced increasing gaps between Atole and Fresco villages to identify the vital structural parameters in our model. 

\subsection{Measurement Error}

We include in the model measurement errors for protein intake and height at age 24 months. The econometrician observes $N^{*}$, which differ from the true optimal nutritional choice $N$ by measurement error $\eta$:
\begin{equation}
N^{*} = N(\Omega_{yv}) \cdot e^{\eta}
\end{equation}

Similarly, we assume that the econometrician observes an error-ridden measure of height at age 24 months, $H^{24,*}$, which differs from actual height by a measurement error $\iota$:

\begin{equation}
H^{24,*} = H^{24}\left(N\left(\Omega_{yv}\right), X, \epsilon\right) \cdot e^{\iota}
\end{equation}

We assume that $\eta$ and $\iota$ are normally distributed, and that $\epsilon$, $\eta$, and $\iota$ are independent. The standard deviation of $\eta$ is $\sigma_{\eta}$ and the mean is $\mu_{\eta}=-\frac{\sigma_{\eta}^2}{2}$. The standard deviation for $\iota$ is $\sigma_{\iota}$ with mean $\mu_{\iota}=-\frac{\sigma_{\iota}^2}{2}$.

\subsection{Identification of Critical Structural Parameters}

For our decomposition exercise, our model requires credible identification of five critical structural parameters: $\mu_{R_{y,v}}$, $\sigma_{R_{y,v}}^{2}$, $\delta$, $\beta$, and $\lambda$. We use equations \ref{eq:mean_belief} to identify $\mu_{R_{y,v}}$. Given the potential presence of both sampling and measurement errors, we cannot use Equation \ref{eq:variance_belief} to identify $\sigma_{R_{y,v}}^{2}$. Rather, we estimate the model given a range of possible $\sigma_{R_{y,v}}^{2}$ values bounded at the lower end under the assumption of no measurement error for the variance parameter from Equation \ref{eq:variance_belief}, and significant measurement error on the upper end. 

Figure \ref{fig:PortHgtGap} shows the gaps in protein intake between Atole and Fresco villages. In our model, the protein-supplementation policy experiment works through the $\delta$ parameter. We estimate $\delta>0$ to match the overall nutritional gap between Atole and Fresco villages.

The parameter $\beta$ captures the impact of protein intake on height production. The challenge of estimating $\beta$ is that protein intake may correlate with the shocks $\epsilon$. We avoid this problem by exploiting information from the experimental study in Guatemala. As we described in Section 3, the experiment induced children in Atole villages to consume more protein. We use the exogenous variation in protein intake from the experiment to pin down the parameter $\beta$. 

Crucial to our model is the reference-point parameter $\lambda$, which guides the impact of reference points on nutritional choices. The identification of this parameter requires exogenous variation in reference points. In our data, such variation is available by exploring information from the interaction of year and exogenous assignment to the treatment group. Figure \ref{fig:PortHgtGap} shows the increasing height gaps between Atole and Fresco villages. This approach is sufficient for identification because there are no statistically significant differences in levels and trends of incomes, prices, gender, and initial heights between Atole and Fresco villages (see our discussion in Section \ref{sec:stat}). 
The model also features other parameters that are not as important for the decomposition exercise. For example, the parameter $\rho$ (the quadratic term of non-child-nutrition consumption $c$) determines the concavity of preferences with respect to $c$. If income does not matter, then the model is quasilinear in income with $\rho=0$. Hence, $\rho$ is identified by the effect of income on choices.

For the linear height preference parameters, if $\gamma=0$ (and $\lambda=0$), that would lead to zero nutritional choices. Given positive nutritional choices, $\gamma>0$. If average nutritional choices are high, $\gamma$ is higher to reflect higher preferences for height and vice-versa.

Finally, the productivity shocks impact both nutritional choices and height outcomes. $\sigma_{\epsilon}$ is identified by the positive covariance between the height at month 24 and nutritional choices that are not captured by income, price, or components of $X$.

In Appendix Section \ref{sec:solu}, we describe how we solve the structural model and present our likelihood function. 

\section{Results}

\subsection{Parameter Estimates\label{sec:param_esti}}

We present estimated parameters in Table \ref{tab:paramestitwo}, with standard errors shown in parentheses. For preferences, $\rho$ is $-0.047$, indicating that preferences are concave in $c$. $\gamma$ and $\lambda$ are $0.033$ and $-0.026$, respectively. Therefore, preferences are also concave in height, and the curvature of the indifference curves indicates the asymmetry in responses. Parents prefer taller to shorter children, but the marginal benefit from an additional centimeter of height beyond the reference comparison height is close to zero (albeit positive).

The price discount parameter $\delta$ is $0.376$, representing a $38\%$ discount in protein prices in Atole villages. The production-function parameters are $\beta=0.073$, $\alpha_{H_0}=0.022$, $\alpha_{male}=0.009$, $A=4.144$, and $\sigma_{\epsilon}=0.010$. Given these parameters, we show the consumption- and height-possibility frontier along with indifference curves for an individual in panel a of Figure \ref{fig:indiff}. 

The measurement-error estimates are $\sigma_{\eta}=0.382$ and $\sigma_{\iota}=0.043$. These findings indicate that measurement error is more severe for nutritional intake than height, which is intuitive.

\subsection{Model Fit\label{sec:model_fit}}

Given estimated parameters, we solve for optimal protein choices for each household. Table \ref{modelfit} compares model and data protein intake and height at month 24 in Fresco and Atole villages. Panel a compares data and model by aggregating across genders and years. Panel b shows the fit of the model by gender, aggregated across years. Panel c shows the fit by years but aggregates over gender. The fit is generally good, but the discrepancy between model prediction and data tends to be greater for protein intake than height. For example, the model’s worst fit is about protein consumption for girls, and the best fit is for girls’ height at age 24 months. Across years, the model generally provides good fits for both protein intake and height, with different outcomes having the worst and the best fit in each year. 

We investigate if the model fits Atole villages’ data better than those from the Fresco counterparts, but we do not find any form of dominance in this comparison. For example, when we break the analysis by gender, it performs worse for boys in Fresco villages and for girls in Atole villages. We conclude that there are no systematic differences in fit by experimental groups. This finding provides validation for the decomposition exercise we perform next. 

\subsection{Decomposition of Atole--Fresco Gap due to Atole Intervention\label{sec:decompose}}

We use the estimated model to decompose the relative contributions of prices and reference points to the height dynamics in Atole and Fresco villages. In Appendix \ref{sec:targetuniversal}, we discuss additional counterfactuals in which we compare the relative impacts of poor-targeted and universal policy experiments given reference points.

The intervention, which started in 1969, leads to a reduction in protein prices in Atole villages. The discount in this price directly and immediately impacts their protein intake, which explains the difference in height between Atole and Fresco villages for the 1970-71 cohort. Therefore, this price reduction is the most significant driving force for the first cohort in our model. The intervention impacted protein intake and height for the remaining cohorts by simultaneously reducing the price and increasing the reference points.

\begin{figure}[H]
    \centering
	\includegraphics[scale=1.0]{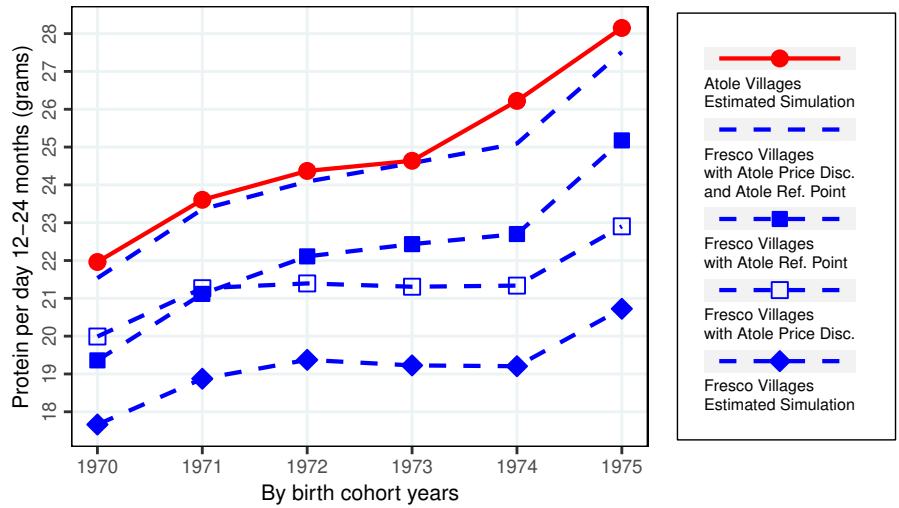}
	\caption{Decompose protein choice gaps between Atole and Fresco villages. \emph{Note}: We evaluate the effects of giving fresco villages Atole price and Atole reference points on the protein choice gaps between Atole and Fresco villages. See Section \ref{sec:decompose} for discussions.}
	\label{fig:frescoatoledecompose}
\end{figure}

\begin{figure}[H]
    \centering
	\begin{subfigure}[t]{1\textwidth}
        \centering
    	\includegraphics[scale=1.0]{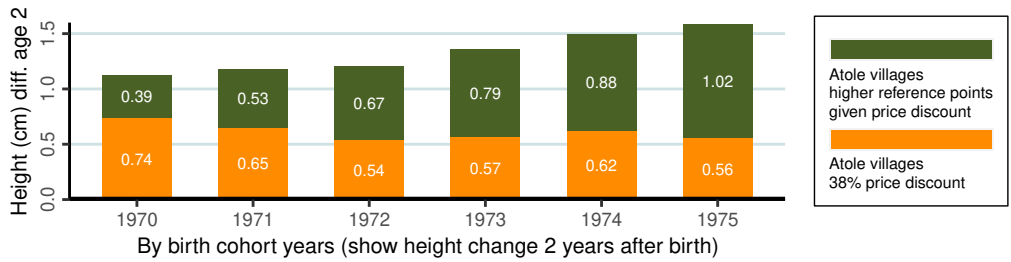}
    	\caption{Level of contributions}
	\end{subfigure}
	\par\smallskip
	\begin{subfigure}[t]{1\textwidth}
	    \centering
    	\includegraphics[scale=1.0]{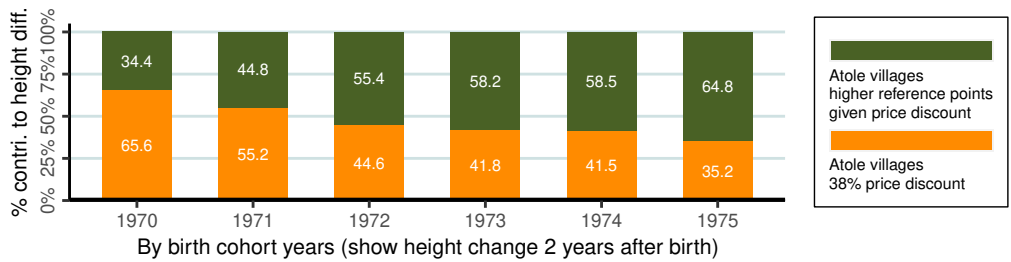}
    	\caption{Share of contributions}
	\end{subfigure}
	\caption{Decompose height gaps between Atole and Fresco villages. \emph{Note}: We compare the contributions, in levels and shares, of reference points changes to the height gaps between Atole and Fresco villages. See Section \ref{sec:decompose} for discussions.}
	\label{fig:frescoatoledecomposehgt}
\end{figure}

Table \ref{frescoatoledecompose} and Figures \ref{fig:frescoatoledecompose} and \ref{fig:frescoatoledecomposehgt} report the results of three counterfactual simulations for Fresco villages that showcase the capacity of our model to separate the contributions of the price reduction from the shifts in reference points. 

In the first column of panel a on Table \ref{frescoatoledecompose}, we summarize the simulated average height outcomes and protein choices for households in Fresco villages. In column five, we show these statistics for households in Atole villages. Comparing these two columns, the height gaps between Atole and Fresco villages grow from 1.19 centimeters in 1970-71 to 1.76 centimeters in 1974-75.\footnote{For the six years of the study, the height gaps between Atole and Fresco villages were 1.12cm, 1.19cm, 1.14cm, 1.41cm, 1.78cm, and 1.79cm.} 

In the second column, we simulate the effect of Fresco households receiving a 38\% reduction in the protein price. This price change reduces the Atole--Fresco height gaps between the first and the fifth columns, by about 0.6 centimeters, with a small cohort-to-cohort variation. 

In column three, we simulate the effect of replacing the reference points in Fresco villages with the Atole reference points without changing prices. The height gaps between the first and the fifth columns decrease by about 0.6 centimeters in 1970-71 (48\%), to almost a full centimeter in 1974-75 (55\%). Figure \ref{fig:frescoatoledecompose} provides a visualization of results in Table \ref{frescoatoledecompose}.

In column four of Table \ref{frescoatoledecompose}, we simulate protein intake and height at age 24 months of children in Fresco villages if the households experienced the price discounts and increment in reference points.\footnote{The resulting outcomes closely approximate the simulated results from Atole villages shown in column five. This result is not surprising given that other state variables do not vary significantly between Atole and Fresco villages, as seen in Table \ref{summcovarmain}.} 

In Figure \ref{fig:frescoatoledecomposehgt}, we decompose the total difference into the effects of the price discounts only (differences between columns two and one), and the additional effects from also imposing reference-points shifts (differences between columns four and two). The price discount's impact is stable at around 0.6 to 0.7 centimeters over time. From 1970 to 1975, the share explained by the price discount decreases from 66\% to 35\% of the total impact. During the same period, the share of the total effects explained by reference points given the price discount increases from 0.39 to 1.02 centimeters, or from 34\% to 65\% of the total impact. 

\subsection{Varying Reference-Points Variance Beliefs \label{sec:varyrefesti}}

In this section, we continue with the counterfactual decomposition analysis. Here, we use alternative estimates based on varying assumptions for $\sigma_{R_{y,v}}^{2}$. Under our main specifications, we assumed that $\sigma_{R_{y,v}}^{2}$ is determined by the variance of the mean belief. It captures the uncertainty parents face with respect to their mean beliefs. However, we do not have direct empirical evidence for the uncertainty in mean beliefs about the reference point. Our ignorance about this parameter matters because the greater the variance beliefs, the less important the reference points, \emph{ceteris paribus}. 

In this section, we explore alternative values for $\sigma_{R_{y,v}}^{2}$. The standard deviation of height at age 24 months is approximately $3.5$ centimeters. The standard error of the mean beliefs is around $0.5$ centimeters. We used $\sigma_{R_{y,v}}=0.5$ for the main specification results discussed so far. \footnote{See discussions and details in Appendix Section \ref{sec:estidata}.} 

\subsubsection{Alternative Model Estimates under Varying $\sigma_{R_{y,v}}$}

To assess the sensitivity of our results, we re-estimate the model for values of $\sigma_{R{y,v}}$ in the set $\{1.5, 2.5, 3.5\}$. We pick $3.5$ as the upper-bound because it is the standard deviation of the height distribution at age 24 months (see our discussion in Appendix Section \ref{sec:appsecalter}). In Table \ref{tab:paramesti}, we present parameter estimates for the model under alternative assumptions on $\sigma_{R_{y,v}}$.

Overall, as we shift assumptions on the value of $\sigma_{R_{y,v}}$, which is a parameter of the utility function, the parameter estimates for preference parameters shift. As $\sigma_{R_{y,v}}$ increases from $0.5$ to $3.5$, estimates for $\rho$ decrease, estimates for $\gamma$ decrease before increasing, and estimates for $\lambda$ largely decrease.\footnote{In Appendix Figure A.2, we visualize the point-estimates for $\rho$, $\gamma$, and $\lambda$ obtained by re-estimating the model along a finer set of $\sigma_{R{y,v}}$ values between 0.5 and 3.5.} 

Given these estimates, at higher $\sigma_{R_{y,v}}$ levels, the marginal gains from additional units of non-child-nutrition consumption decrease. At the $\sigma_{R_{y,v}}=3.5$, $0.035 = \gamma < |\lambda| = |-0.041|$, hence the marginal gains from an additional unit of height are slightly decreasing beyond the mean of the reference points distribution. This is true for $\sigma_{R_{y,v}}=2.5$ as well. At $\sigma_{R_{y,v}}=0.5$, $0.033 = \gamma > |\lambda| = |-0.026|$, hence the expected marginal gains from an additional unit of height are slightly increasing beyond the mean of the reference-points distribution. This is true for $\sigma_{R_{y,v}}=1.5$ as well. In essence, the higher the value of $\sigma_{R_{y,v}}$, the higher (in absolute value) $\lambda$. Therefore, while greater values of $\sigma_{R_{y,v}}$ reduce the importance of reference points, the greater (absolute) values of $\lambda$ make reference points more relevant. This finding will explain why our decomposition exercise is largely invariant to $\sigma_{R_{y,v}}$.

Production function, price discount, and measurement error parameters are generally stable across assumptions for the values for $\sigma_{R_{y,v}}$. As preference parameters adjust to match the same data on nutritional choices, it does not surprise us that the parameters for the production function and budget are relatively the same: the input and output relationship has not changed, and the relative gaps in height outcomes and nutritional choices across Atole and Fresco villages across cohorts have not changed. The measurement error distribution is essentially the same across values for $\sigma_{R_{y,v}}$ as well because $\sigma_{R_{y,v}}$ is a parameter that shifts the curvature of indifference curves. Higher $\sigma_{R_{y,v}}$ does not expand the variance of observed shocks facing households and has limited impacts on the variance of model predicted choices and expected heights.\footnote{In Appendix Table \ref{tab:paramfitmulti}, we show that the fits of the model are very similar under different $\sigma_{R_{y,v}}$ assumptions. Given our data, it is not possible for us to identify $\sigma_{R_{y,v}}$ directly from nutritional choices and height outcome patterns across cohorts. Preference parameters are able to adjust flexibly to accommodate varying $\sigma_{R_{y,v}}$ values and provide similar model fits. This non-identification result is not new in the literature that allows for agents to have heterogeneous beliefs, as described in \textcite{manski_2004}.}

\subsubsection{Decomposition under Varying $\sigma_{R_{y,v}}$}

\begin{figure}[H]
    \centering
	\begin{subfigure}[t]{1\textwidth}
        \centering
    	\includegraphics[scale=1.0]{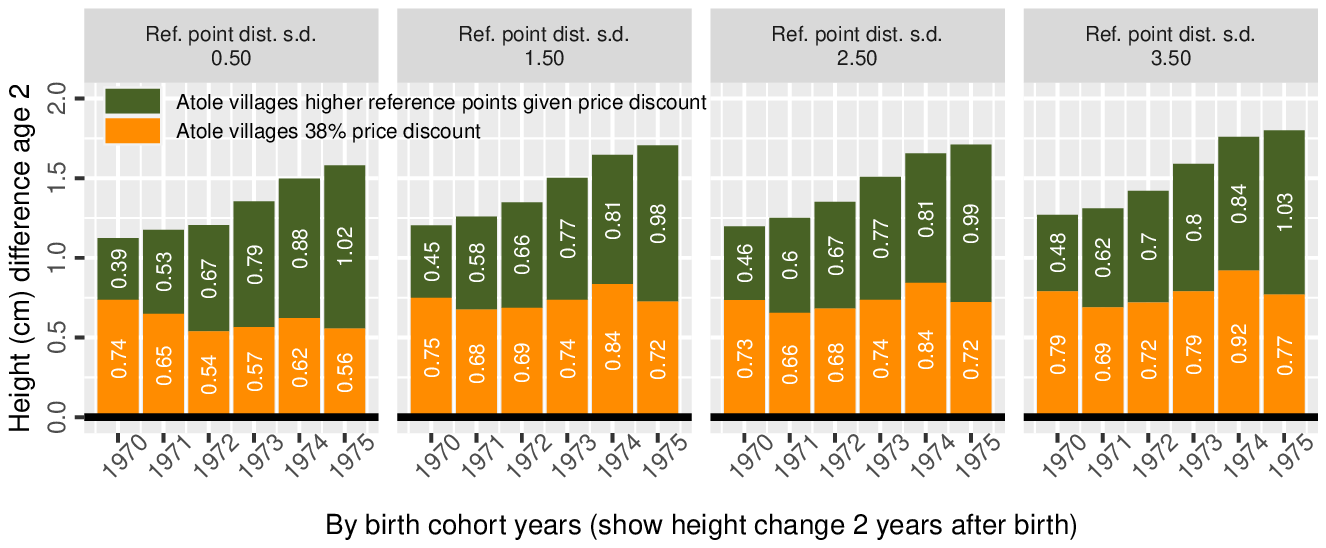}
    	\caption{Level of contributions}
	\end{subfigure}
	\par\smallskip
	\begin{subfigure}[t]{1\textwidth}
	    \centering
    	\includegraphics[scale=1.0]{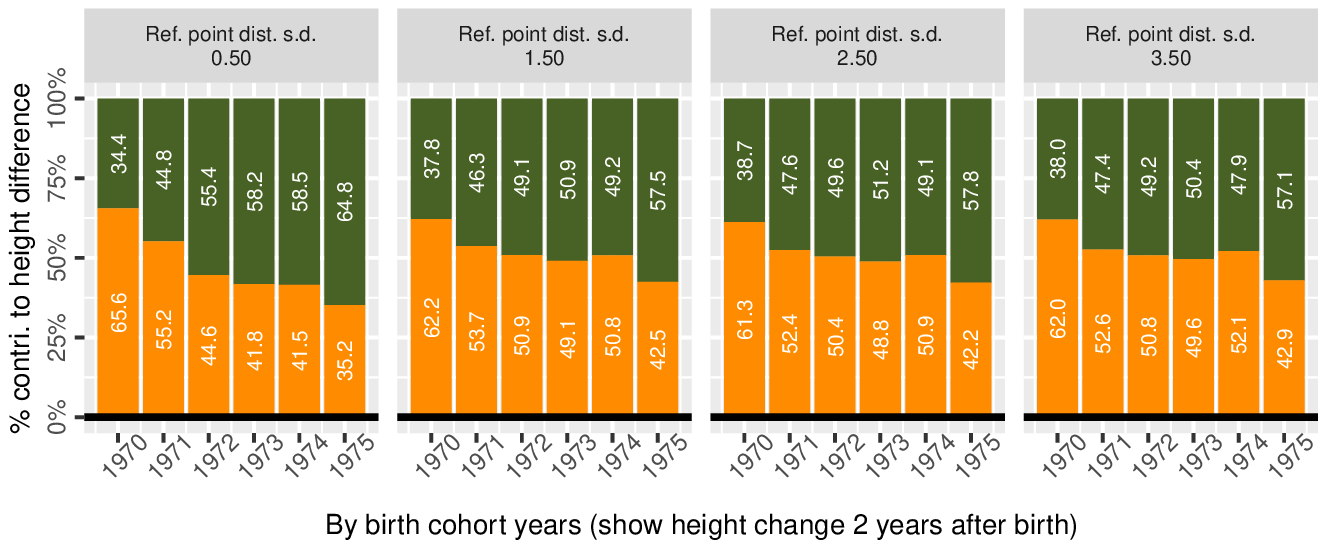}
	    \caption{Share of contributions}
	\end{subfigure}
	\caption{Decompose height gaps between Atole and Fresco villages with varying $\sigma_R$.
	\emph{Note}: We re-estimate the model under different reference points standard deviation assumptions. Given estimates at selected $\sigma_R$ values, we compare the contributions, in levels and shares, of reference points changes to the height gaps between Atole and Fresco villages. See Section \ref{sec:varyrefesti} for discussions.}
	\label{fig:frescoatoledecomposemultiple}
\end{figure}

Given the different parameter estimates as reported in Table \ref{tab:paramesti}, we conduct the counterfactual decomposition exercises for $\sigma_{R_{y,v}}$ between 0.5 and 3.5~cm. Figure \ref{fig:frescoatoledecomposemultiple} provides graphical illustrations of the decompositional results under alternative assumptions on $\sigma_{R_{y,v}}$. Similar to Figure  \ref{fig:frescoatoledecomposehgt}, the results compare the counterfactual scenarios: changing protein price; changing both prices and reference points simultaneously.

As shown in panel a of \ref{fig:frescoatoledecomposemultiple}, the price discount's impacts are stable and vary between 0.6 and 0.8 centimeters across cohorts and across $\sigma_{R_{y,v}}$ assumptions. The results from $\sigma_{R_{y,v}} = 0.5$ show a decreasing height effect attributed to the price discount, however for larger $\sigma_{R_{y,v}}$ the price effect remains fairly constant across cohorts. Since the price discount is common across the years, this indicates that at the estimated parameter for $\sigma_{R_{y,v}}=0.5$, household choices are more sensitive to variations in prices and income. The impact of reference-point increases, given the price discount, is monotonically rising from the 1970 to the 1975 cohorts under all assumptions for $\sigma_{R_{y,v}}$. For the 1970 cohort, the impact varies between 0.4 to 0.5~cm across $\sigma_{R_{y,v}}$. For the 1973 cohort, the impacts are stable at around 0.80 centimeters. For the 1975 cohort, the impacts vary just a little around one centimeter.

In percentage terms, from 1970 to 1975, the contributions of the price discount decrease under all assumptions for $\sigma_{R_{y,v}}$. This is shown in panel b of \ref{fig:frescoatoledecomposemultiple}. We can use the different assumptions on $\sigma_{R_{y,v}}$ to bound the share of the impact of the intervention due to changing protein prices. For example, for the cohort of children born in 1975, the change in the protein price explains 35\% to 43\% of the impact on the height at age 24 months. The change in reference points, given changes in the price, represents 57\% to 65\% of the total impact. In short, our sensitivity analysis shows that for the cohorts born in 1973 or later, the change in the reference point is the primary driver of the change in height at age 24 months.

\section{Conclusion}

In this paper, we build and estimate a child-nutritional investment model. The model considers reference-dependent preferences, where the reference is with respect to the heights of the previous cohort of children who live in the same village. Reference points shift endogenously as households observe children’s heights from earlier birth cohorts changing.

For researchers interested in the impact of price subsidies and income transfers, we have introduced a long-term secondary channel -- endogenous changes in reference points -- that might affect the impacts of these policies. For the protein-supplement experiment implemented in Guatemala, which we interpret as a price-discount policy, by 1975 -- six years after the start of the policy – 57\% to 65\% of the policy’s impact are due to its impact on shifting reference points.

Our paper also shows significant height increases might be realized from shifting reference points for highly-stunted populations. It is an open question how to exogenously shift these reference points in the short run, although the Peruvian experience mentioned in the introduction indicates that educational campaigns could be effective on a large scale over time. The cost of an educational campaign to inform households about alternative reference heights might be lower than income transfers and price subsidies with similar effects on heights.

We recognize that our analysis requires a crucial assumption: parents use the data on a selected group of children to estimate mean and variance beliefs about the reference height. We argued that our assumption is consistent with research in economics, medicine, and anthropology, but we recognize that there are many other alternatives, such as rational expectations, Bayesian updating, or social learning models. Unfortunately, the literature in economics knows very little about such a critical element of the model we propose in this paper. Therefore, it is necessary to encourage research that elicits parental subjective distributions of reference points as part of parent-directed interventions to reduce stunting. Within a randomized controlled trial, the availability of such data would shed light on the process parents use to update critical moments of the subjective distribution. With such evidence, one could have a more robust representation of our model that would better inform the design of public policies to improve children’s health. 

\clearpage
\pagebreak

\begingroup
\setstretch{1.0}
\setlength\bibitemsep{3pt}
\printbibliography[title=References]
\endgroup
\pagebreak


\newcommand{\subFigWidth}{0.50}
\begin{figure}[H]
    \centering
    \begin{subfigure}[b]{\subFigWidth\textwidth}
        \centering
        \includegraphics[width=\linewidth]{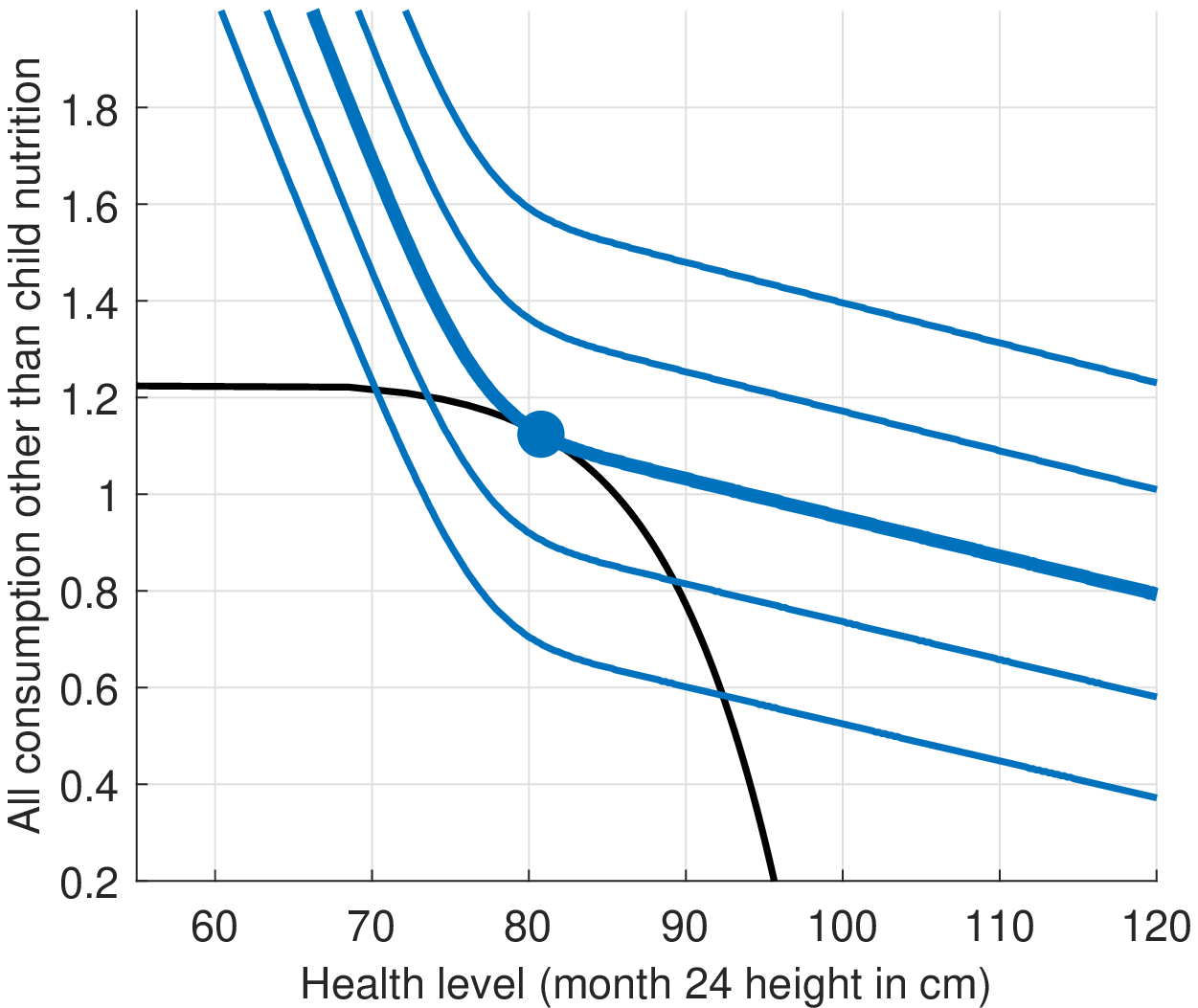}
        \caption{Contours of indifference curves at given estimates}
    \end{subfigure}~
	\begin{subfigure}[b]{\subFigWidth\textwidth}
	    \centering
    	\includegraphics[width=\linewidth]{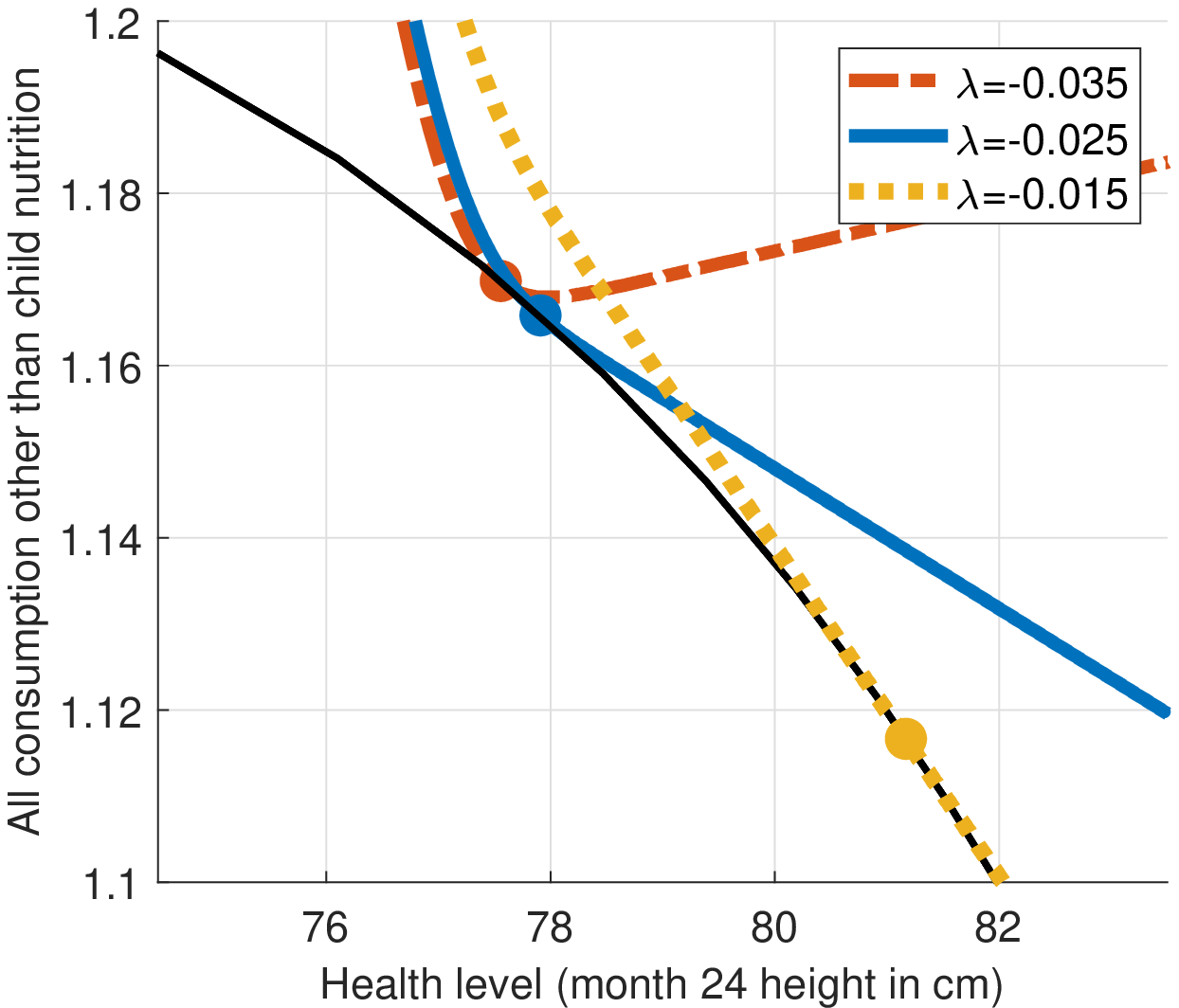}
    	\caption{Vary $\lambda$}
	\end{subfigure}
	\par\medskip
	\begin{subfigure}[b]{\subFigWidth\textwidth}
        \centering
        \includegraphics[width=\linewidth]{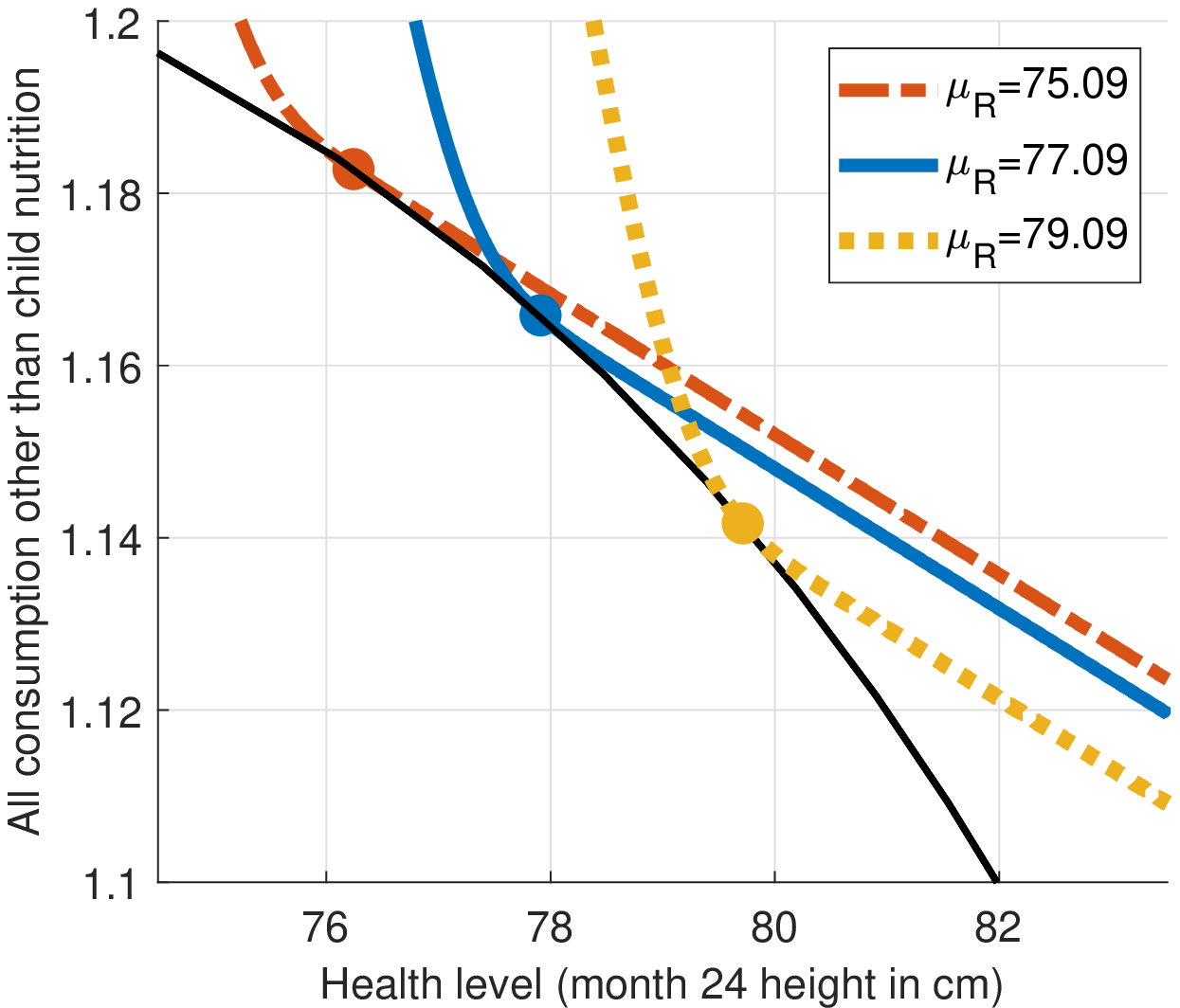}
        \caption{Vary $\mu_{R}$}
    \end{subfigure}~
	\begin{subfigure}[b]{\subFigWidth\textwidth}
	    \centering
	    \includegraphics[width=\linewidth]{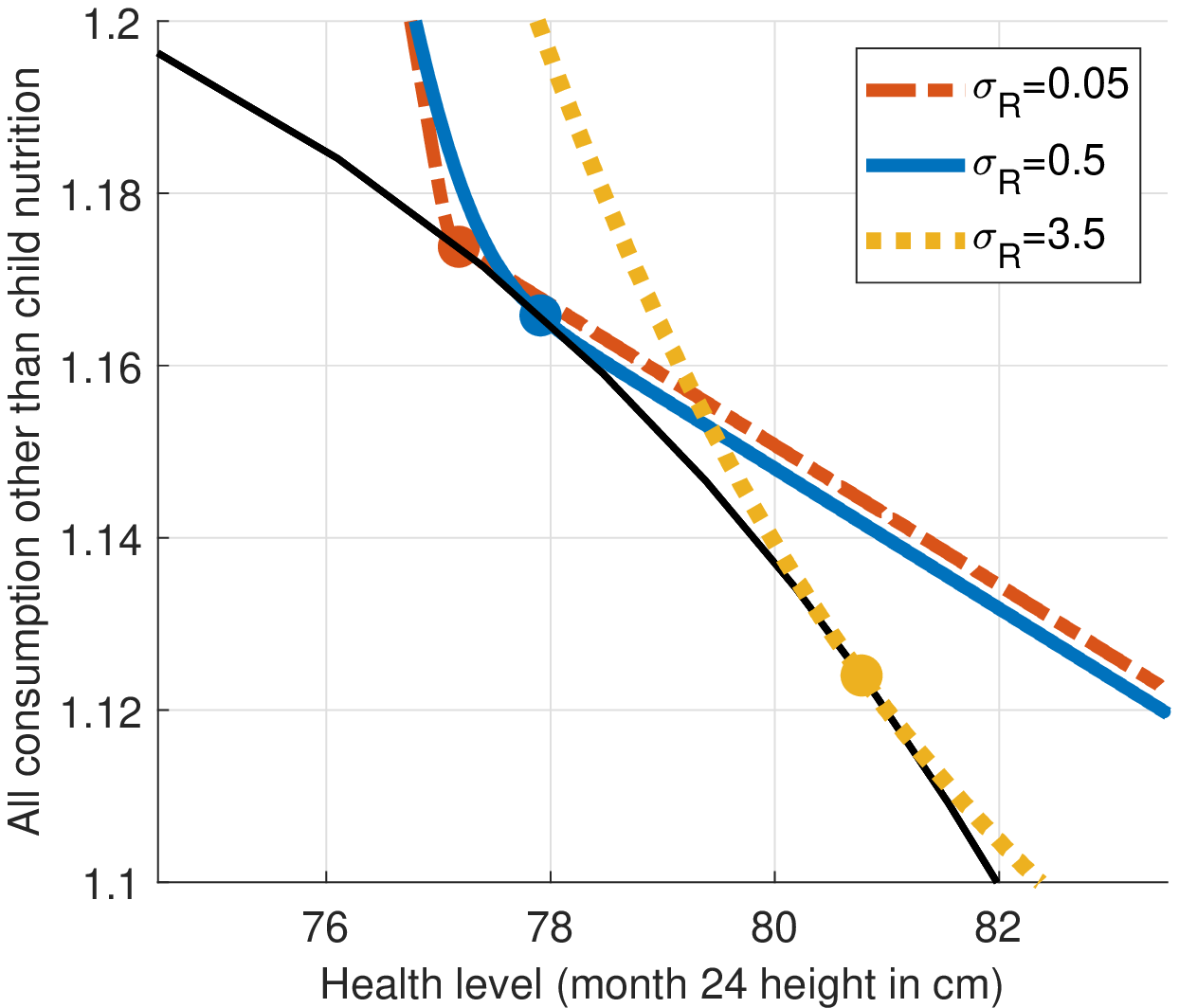}
	    \caption{Vary $\sigma_{R}$}
	\end{subfigure}
	\caption{Consumption and height outcome indifference curves and choice frontiers. \emph{Note}: Health possibility frontier and indifference curves are visualized in blue (solid-line) given estimated parameters and characteristics for one particular child. The consumption and expected height outcome frontier is determined by the budget constraint as well as the production function. Panels b, c and d show results from varying one parameter at a time in red (dashed-line) and orange (dotted-line). See Section \ref{sec:maximize} for discussions.}
	\label{fig:indiff}
\end{figure}
\pagebreak

\begin{figure}[H]
    \centering
    \begin{subfigure}[t]{.43\textwidth}
        \centering
        \includegraphics[width=\linewidth]{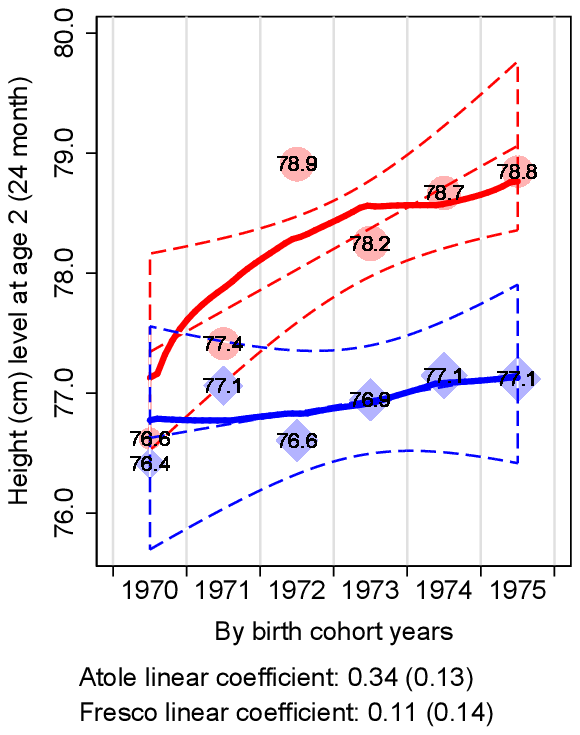}
        \caption{Height}
    \end{subfigure}~
	\begin{subfigure}[t]{0.57\textwidth}
	    \centering
    	\includegraphics[width=\linewidth]{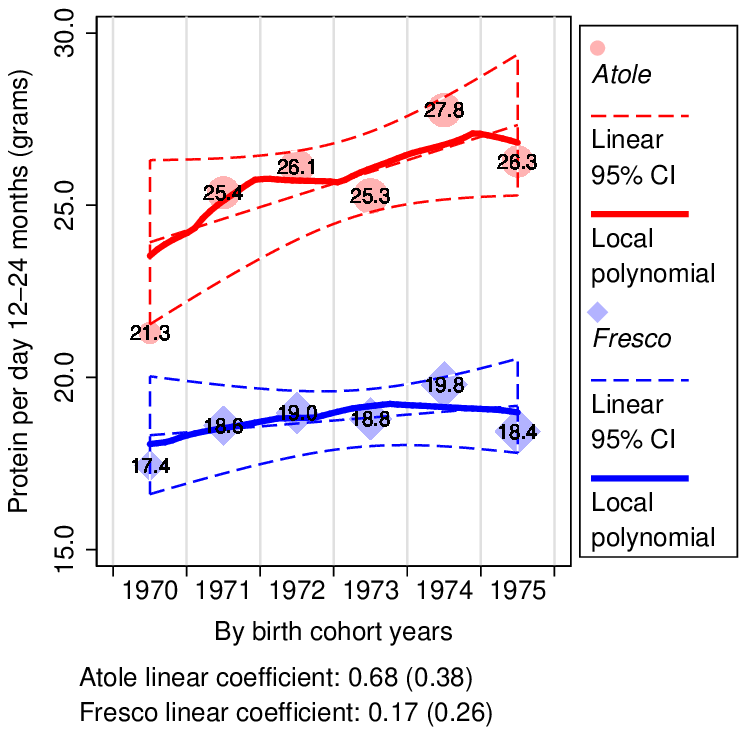}
    	\caption{Protein}
	\end{subfigure}
	\caption{Increasing protein and height gaps across cohorts. \emph{Note}: See Section \ref{sec:gap} for discussions.}
	\label{fig:PortHgtGap}
\end{figure}
\pagebreak

\pagebreak
\pagebreak

\pagebreak

\pagebreak

\clearpage

\begin{table}[t!]
{\small
\centering                                 \def\sym#1{\ifmmode^{#1}\else\(^{#1}\)\fi}                                 \caption{Summary statistics for various variables\label{summcovarmain}}                                 \begin{adjustbox}{max width=1\textwidth}                                 \begin{tabular}{m{6.9cm} >{\centering\arraybackslash}m{1.8cm} >{\centering\arraybackslash}m{1.8cm} >{\centering\arraybackslash}m{1.8cm} >{\centering\arraybackslash}m{1.8cm} >{\centering\arraybackslash}m{1.8cm}}                                 \toprule                                                                                         & \multicolumn{5}{C{9cm}}{Atole and Fresco villages differences} \\                                                 \cmidrule(l{5pt}r{5pt}){2-6}                                                   & \multicolumn{1}{C{1.8cm}}{\small \textbf{All}} & \multicolumn{2}{C{3.6cm}}{\small \textbf{Group averages}} & \multicolumn{2}{C{3.6cm}}{\small \textbf{p-values testing}} \\                                                  \cmidrule(l{5pt}r{5pt}){2-2} \cmidrule(l{5pt}r{5pt}){3-4} \cmidrule(l{5pt}r{5pt}){5-6}                                                  & \multicolumn{1}{C{1.8cm}}{\textit{\small mean (sd)}} & \multicolumn{1}{C{1.8cm}}{\textit{\small Fresco}} & \multicolumn{1}{C{1.8cm}}{\textit{\small Atole}} & \multicolumn{1}{C{1.8cm}}{\textit{\small gap}} & \multicolumn{1}{C{1.8cm}}{\textit{\small p-value}} \\
\midrule
\multicolumn{6}{l}{\vspace*{-3mm}\hspace*{-0mm}\textbf{\normalsize Panel a: Gender income price (N=503, main sample)}} \\                                          &            &            &            &            &            \\
Male                &        0.52&        0.52&        0.52&       -0.00&        0.92\\
                    &\vspace*{-2mm}{\footnotesize (0.50) }&\vspace*{-2mm}{\footnotesize (0.50) }&\vspace*{-2mm}{\footnotesize (0.50) }&            &            \\
Income (quetzale)       &      515.57&      503.68&      526.00&       22.32&        0.59\\
                    &\vspace*{-2mm}{\footnotesize (460.9) }&\vspace*{-2mm}{\footnotesize (464.4) }&\vspace*{-2mm}{\footnotesize (458.4) }&            &            \\
Mth 15-24 protein price (quetzale/10k grams)&       52.58&       52.47&       52.68&        0.21&        0.54\\
                    &\vspace*{-2mm}{\footnotesize (3.87) }&\vspace*{-2mm}{\footnotesize (3.93) }&\vspace*{-2mm}{\footnotesize (3.81) }&            &            \\
\midrule
\multicolumn{6}{l}{\vspace*{-3mm}\hspace*{-0mm}\textbf{\normalsize Panel b: Gender income (N=1115, height observed once in first 24 months)}} \\                                          &            &            &            &            &            \\
Male                &        0.53&        0.53&        0.53&        0.00&        0.98\\
                    &\vspace*{-2mm}{\footnotesize (0.50) }&\vspace*{-2mm}{\footnotesize (0.50) }&\vspace*{-2mm}{\footnotesize (0.50) }&            &            \\
Income (quetzale)       &      449.49&      444.63&      454.06&        9.43&        0.72\\
                    &\vspace*{-2mm}{\footnotesize (432.3) }&\vspace*{-2mm}{\footnotesize (446.4) }&\vspace*{-2mm}{\footnotesize (419.0) }&            &            \\
\midrule
\multicolumn{6}{l}{\vspace*{-3mm}\hspace*{-0mm}\textbf{\normalsize Panel c: Height}} \\                                          &            &            &            &            &            \\
Month 0 (cm)  {\footnotesize N=503}&       49.64&       49.79&       49.52&       -0.27&        0.19\\
                    &\vspace*{-2mm}{\footnotesize (2.29) }&\vspace*{-2mm}{\footnotesize (2.29) }&\vspace*{-2mm}{\footnotesize (2.29) }&            &            \\
Month 6 (cm)  {\footnotesize N=463}&       62.72&       62.49&       62.93&        0.44&        0.05\\
                    &\vspace*{-2mm}{\footnotesize (2.46) }&\vspace*{-2mm}{\footnotesize (2.50) }&\vspace*{-2mm}{\footnotesize (2.42) }&            &            \\
Month 12 (cm) {\footnotesize N=475}&       68.81&       68.45&       69.13&        0.68&        0.01\\
                    &\vspace*{-2mm}{\footnotesize (2.99) }&\vspace*{-2mm}{\footnotesize (3.13) }&\vspace*{-2mm}{\footnotesize (2.83) }&            &            \\
Month 18 (cm) {\footnotesize N=482}&       73.37&       72.88&       73.80&        0.92&        0.00\\
                    &\vspace*{-2mm}{\footnotesize (3.23) }&\vspace*{-2mm}{\footnotesize (3.26) }&\vspace*{-2mm}{\footnotesize (3.15) }&            &            \\
Month 24 (cm) {\footnotesize N=503}&       77.66&       76.97&       78.30&        1.33&        0.00\\
                    &\vspace*{-2mm}{\footnotesize (3.47) }&\vspace*{-2mm}{\footnotesize (3.49) }&\vspace*{-2mm}{\footnotesize (3.33) }&            &            \\
\midrule
\multicolumn{6}{l}{\vspace*{-3mm}\hspace*{-0mm}\textbf{\normalsize Panel d: Average daily nutritional intake}} \\                                          &            &            &            &            &            \\
Month 15 protein (grams/day) {\footnotesize N=464}&       17.43&       14.29&       20.07&        5.78&        0.00\\
                    &\vspace*{-2mm}{\footnotesize (10.5) }&\vspace*{-2mm}{\footnotesize (9.14) }&\vspace*{-2mm}{\footnotesize (10.9) }&            &            \\
Month 18 protein (grams/day) {\footnotesize N=461}&       21.52&       18.27&       24.41&        6.14&        0.00\\
                    &\vspace*{-2mm}{\footnotesize (11.4) }&\vspace*{-2mm}{\footnotesize (9.61) }&\vspace*{-2mm}{\footnotesize (12.0) }&            &            \\
Month 21 protein (grams/day) {\footnotesize N=475}&       24.45&       20.17&       27.99&        7.82&        0.00\\
                    &\vspace*{-2mm}{\footnotesize (11.4) }&\vspace*{-2mm}{\footnotesize (9.03) }&\vspace*{-2mm}{\footnotesize (11.9) }&            &            \\
Month 24 protein (grams/day) {\footnotesize N=462}&       26.99&       22.51&       31.07&        8.56&        0.00\\
                    &\vspace*{-2mm}{\footnotesize (12.0) }&\vspace*{-2mm}{\footnotesize (9.02) }&\vspace*{-2mm}{\footnotesize (13.0) }&            &            \\
Avg mth 15-24 protein (grams/day) {\footnotesize N=503}&       22.54&       18.78&       25.84&        7.06&        0.00\\
                    &\vspace*{-2mm}{\footnotesize (8.98) }&\vspace*{-2mm}{\footnotesize (6.43) }&\vspace*{-2mm}{\footnotesize (9.62) }&            &            \\
Avg mth 15-24 non-protein (kcal/day)   {\footnotesize N=503}&      691.78&      681.78&      700.55&       18.77&        0.38\\
                    &\vspace*{-2mm}{\footnotesize (236.5) }&\vspace*{-2mm}{\footnotesize (236.2) }&\vspace*{-2mm}{\footnotesize (236.9) }&            &            \\
\bottomrule
\addlinespace[0.5em]
\multicolumn{6}{p{1\textwidth}}{\parbox[t]{1.0\textwidth}{\footnotesize{\emph{Note}: See Section \ref{sec:stat} for discussions.}}}\\
\addlinespace
\end{tabular}
\end{adjustbox}
}
\end{table}

\clearpage
\begin{table}[!ht]
\centering{}
{\small
	\caption{\label{tab:paramestitwo}Estimated model parameters ($\sigma_R =0.50$ cm)}
	\begin{tabular}{crlcrl}
		\toprule
		\multicolumn{6}{c}{\textbf{Parameter estimates (s.e.) }} \\
		\midrule
		\multicolumn{6}{c}{\vspace{-3mm}} \\
		\multicolumn{3}{l}{\textit{Preference}}	& \multicolumn{3}{l}{\textit{Production function}} \\
		\multicolumn{6}{c}{\vspace{-3mm}} \\
		$\rho$  & $-0.0473$  & $(0.0031) $ 	 	&        $A$ & $4.1435$ & $(0.0854)$ \\
		$\gamma$ & $0.0325$  & $(0.0059) $ 	    &    	 $\alpha_{H_0}$ & $0.0220$ & $(0.0088)$ \\
		$\lambda$ & $-0.0257$ & $(0.0056) $      &        $\alpha_{male}$ & $0.0086$ & $(0.0016)$ \\
   		\multicolumn{2}{c}{} 	              & &        $\beta$ & $0.0725$ & $(0.0158)$  \\
   		\multicolumn{2}{c}{} 	              & &        $\sigma_\epsilon$ & $0.0097$ &   $(0.0010)$ \\
		\multicolumn{6}{c}{} \\
		\multicolumn{3}{l}{\textit{Price discount}} & \multicolumn{3}{l}{\textit{Measurement error}} \\
		\multicolumn{6}{c}{\vspace{-3mm}} \\
		$\delta$ & $0.3756$ & $(0.0249)$ & $\sigma_\eta$ & $0.3823$ & $(0.0151)$ \\
   		\multicolumn{2}{c}{} 	              & &        $\sigma_\iota$ & $0.0427$ &   $(0.0017)$ \\
		\multicolumn{6}{c}{\vspace{-3mm}} \\
        \bottomrule
        \addlinespace[0.5em]
        \multicolumn{6}{l}{\footnotesize{\emph{Note}: See Section \ref{sec:param_esti} for discussions.}}\\
	\end{tabular}
}
\end{table}
\clearpage
\newcommand{\sefm}[1]{\vspace*{-2mm}{\footnotesize (#1)}}
\begin{table}[htbp]
\centering
{\small
\def\sym#1{\ifmmode^{#1}\else\(^{#1}\)\fi}
\caption{The fit of the estimated model's simulated choices to data\label{modelfit}}
\begin{adjustbox}{max width=1\textwidth}
\begin{tabular}{m{1.3cm}
>{\centering\arraybackslash}m{0.9cm} 
>{\centering\arraybackslash}m{0.9cm} 
>{\centering\arraybackslash}m{0.9cm} 
>{\centering\arraybackslash}m{0.9cm}
>{\centering\arraybackslash}m{0.9cm}
>{\centering\arraybackslash}m{0.9cm}
>{\centering\arraybackslash}m{0.9cm}
>{\centering\arraybackslash}m{0.9cm}
>{\centering\arraybackslash}m{0.9cm}
>{\centering\arraybackslash}m{0.9cm}
>{\centering\arraybackslash}m{0.9cm}
>{\centering\arraybackslash}m{0.9cm}}
\toprule
& \multicolumn{6}{C{5.4cm}}{\hspace*{-10mm}\textbf{Average protein choices (s.e)}} 
& \multicolumn{6}{C{5.4cm}}{\hspace*{-15mm}\textbf{Average height outcome (s.e)}} \\
\cmidrule(l{5pt}r{5pt}){2-7} \cmidrule(l{5pt}r{5pt}){8-13}
& 
\multicolumn{3}{C{2.7cm}}{\small \textbf{Fresco}} & 
\multicolumn{3}{C{2.7cm}}{\small \textbf{Atole}} & 
\multicolumn{3}{C{2.7cm}}{\small \textbf{Fresco}} & 
\multicolumn{3}{C{2.7cm}}{\small \textbf{Atole}} \\
\cmidrule(l{5pt}r{5pt}){2-4} \cmidrule(l{5pt}r{5pt}){5-7} \cmidrule(l{5pt}r{5pt}){8-10} \cmidrule(l{5pt}r{5pt}){11-13}
& 
\multicolumn{1}{C{0.9cm}}{\textit{\footnotesize{Model}}} & 
\multicolumn{1}{C{0.9cm}}{\textit{\footnotesize{Data}}} &
\multicolumn{1}{C{0.9cm}}{\textit{\footnotesize{p-val}}} & 
\multicolumn{1}{C{0.9cm}}{\textit{\footnotesize{Model}}} & 
\multicolumn{1}{C{0.9cm}}{\textit{\footnotesize{Data}}} & 
\multicolumn{1}{C{0.9cm}}{\textit{\footnotesize{p-val}}} & 
\multicolumn{1}{C{0.9cm}}{\textit{\footnotesize{Model}}} & 
\multicolumn{1}{C{0.9cm}}{\textit{\footnotesize{Data}}} & 
\multicolumn{1}{C{0.9cm}}{\textit{\footnotesize{p-val}}} & 
\multicolumn{1}{C{0.9cm}}{\textit{\footnotesize{Model}}} & 
\multicolumn{1}{C{0.9cm}}{\textit{\footnotesize{Data}}} & 
\multicolumn{1}{C{0.9cm}}{\textit{\footnotesize{p-val}}}
\\
\midrule
\multicolumn{13}{L{16.5cm}}{\vspace*{-5mm}\hspace*{-8mm}\textbf{\normalsize Panel a: Average}} \\
             &            &            &     &           &            &      &           &            &      &           &            &\\
All          &       19.38&       18.78& 0.18&      25.10&       25.84& 0.24 &      76.77&       76.97& 0.44 &      78.19&       78.30& 0.52\\
             & \sefm{0.45}& \sefm{0.42}&     &\sefm{0.64}& \sefm{0.59}&      &\sefm{0.26}& \sefm{0.23}&      &\sefm{0.17}& \sefm{0.20}&\\
\midrule
\multicolumn{13}{L{16.5cm}}{\vspace*{-5mm}\hspace*{-8mm}\textbf{\normalsize Panel b: Averages across genders}} \\
                    &            &            & &           &            & &           &            & &           &             &\\
Female              &       18.15&       17.31& 0.18&      23.96&       25.19& 0.13&      76.07&       76.19& 0.75 &      77.60&       77.69 & 0.72\\
                    & \sefm{0.63}& \sefm{0.61}& &\sefm{0.80}& \sefm{0.81}& &\sefm{0.37}& \sefm{0.30}& &\sefm{0.25}& \sefm{0.25} &\\
Male                &       20.48&       20.10& 0.52&      26.16&       26.43& 0.81&      77.40&       77.67& 0.36&      78.74&       78.86 & 0.63\\
                    & \sefm{0.60}& \sefm{0.56}& &\sefm{1.11}& \sefm{0.85}& &\sefm{0.30}& \sefm{0.33}& &\sefm{0.25}& \sefm{0.30} &\\
\midrule
\multicolumn{13}{L{16.5cm}}{\vspace*{-5mm}\hspace*{-8mm}\textbf{\normalsize Panel c: Averages across years}}\\
                    &            &            & &            &            & &           &            & &           &             &\\
1970-71             &       18.43&       18.17& 0.78&       23.13&       24.19& 0.37&      76.52&       76.83& 0.47&      77.71&       77.18 &0.18\\
                    & \sefm{0.95}& \sefm{0.80}& & \sefm{1.18}& \sefm{1.12}& &\sefm{0.43}& \sefm{0.41}& &\sefm{0.39}& \sefm{0.40} &\\
1972-73             &       19.30&       18.97& 0.68&       24.51&       25.68& 0.21 &      76.80&       76.73& 0.79&      78.07&       78.57&0.13\\
                    & \sefm{0.79}& \sefm{0.74}& & \sefm{0.93}& \sefm{0.88}& &\sefm{0.27}& \sefm{0.42}& &\sefm{0.33}& \sefm{0.32} &\\
1974-75             &       19.97&       18.96& 0.17&       27.13&       27.15& 0.99&      76.89&       77.24& 0.29&      78.65&       78.77 & 0.70\\
                    & \sefm{0.74}& \sefm{0.66}& & \sefm{1.14}& \sefm{1.06}& &\sefm{0.33}& \sefm{0.36}& &\sefm{0.31}& \sefm{0.33} &\\
\bottomrule
\addlinespace[0.5em]
\multicolumn{13}{p{1\textwidth}}{\parbox[t]{1.07\textwidth}{\footnotesize{\emph{Note}: For model columns, we report the model population mean predictions. Standard errors in model columns are based on the standard deviation of simulated sample means, with each simulated sample having the same sample size as the observed data. For data columns, we report sub-group observed sample means. Standard errors are based on the observed sample standard deviation and sample size. The p-value column reports the probability of results that deviate further from the observed sample mean occurring, given the sampling distribution given estimated parameters. See Section \ref{sec:model_fit} for discussions.}}}\\
\addlinespace
\end{tabular}\end{adjustbox}
}
\end{table}

\clearpage
\begin{table}[htbp]
\centering                                 \def\sym#1{\ifmmode^{#1}\else\(^{#1}\)\fi}                                 \caption{Decompose the protein gap between Atole and Fresco villages\label{frescoatoledecompose}}                                 \begin{adjustbox}{max width=1\textwidth}                                 \begin{tabular}{m{2.5cm} >{\centering\arraybackslash}m{1.8cm} >{\centering\arraybackslash}m{1.8cm} >{\centering\arraybackslash}m{1.8cm} >{\centering\arraybackslash}m{1.8cm} >{\centering\arraybackslash}m{1.8cm}}                                 \toprule
& \multicolumn{5}{C{9cm}}{Average protein choice across birth years} \\
\cmidrule(l{5pt}r{5pt}){2-6}
& \multicolumn{1}{C{1.8cm}}{\small \textbf{Fresco}} & \multicolumn{3}{C{5.4cm}}{\small \textbf{Fresco counterfactuals}} & \multicolumn{1}{C{1.8cm}}{\small \textbf{Atole}} \\
\cmidrule(l{5pt}r{5pt}){2-2} \cmidrule(l{5pt}r{5pt}){3-5} \cmidrule(l{5pt}r{5pt}){6-6}
& \multicolumn{1}{C{1.8cm}}{\textit{\small simulated without counterfactuals}} & \multicolumn{1}{C{1.8cm}}{\textit{\small Fresco with Atole price discount}} & \multicolumn{1}{C{1.8cm}}{\textit{\small Fresco with Atole ref. point}} & \multicolumn{1}{C{1.8cm}}{\textit{\small Fresco with both Atole price and ref. point}} & \multicolumn{1}{C{1.8cm}}{\textit{\small simulated without counterfactuals}} \\                                         
\midrule
\multicolumn{6}{L{13.3cm}}{\vspace*{-5mm}\hspace*{-5mm}\textbf{\normalsize Panel a: Height across birth cohorts}} \\
&            &            &            &            &            \\
1970-71             &       76.52&       77.20&       77.09&       77.68&       77.71\\
1972-73             &       76.80&       77.35&       77.57&       78.08&       78.07\\
1974-75             &       76.89&       77.48&       77.86&       78.43&       78.65\\
\midrule
\multicolumn{6}{L{13.3cm}}{\vspace*{-5mm}\hspace*{-5mm}\textbf{\normalsize Panel b: Protein across birth cohorts}} \\                                          &            &            &            &            &            \\
1970-71             &       18.43&       20.80&       20.48&       22.69&       23.13\\
1972-73             &       19.30&       21.35&       22.27&       24.33&       24.51\\
1974-75             &       19.97&       22.13&       23.95&       26.32&       27.13\\
\bottomrule
\addlinespace[0.5em]
\multicolumn{6}{l}{\footnotesize{\emph{Note}: See Section \ref{sec:decompose} for discussions.}}\\
\addlinespace
\end{tabular}
\end{adjustbox}
\end{table}
\clearpage
\begin{table}[!ht]
\renewcommand{\arraystretch}{1.2}
\centering
\caption{\hspace*{0mm}\label{tab:paramesti}Parameter estimates from estimating the model under different assumptions about the standard deviation of the reference-points distribution $\sigma_{R}$}
\begin{adjustbox}{max width=1\textwidth}
\begin{tabular}{lcccccccc}
\toprule
& \multicolumn{8}{c}{Parameter estimates (s.e.)}\\
\cmidrule(l{5pt}r{5pt}){2-9}
& \multicolumn{2}{c}{$\sigma_R$ = 0.5 cm} & \multicolumn{2}{c}{$\sigma_R$ =1.5 cm} & \multicolumn{2}{c}{$\sigma_R$ =2.5 cm} & \multicolumn{2}{c}{$\sigma_R$ =3.5 cm}\\
\cmidrule(l{5pt}r{5pt}){2-3} \cmidrule(l{5pt}r{5pt}){4-5} \cmidrule(l{5pt}r{5pt}){6-7} \cmidrule(l{5pt}r{5pt}){8-9}
\multicolumn{9}{l}{\hspace*{0mm}Preference and price discount}\\
\addlinespace
\hspace*{6mm}$\rho$ & -0.0473 & (0.0031) & -0.0662 & (0.0034) & -0.0712 & (0.0044) & -0.0725 & (0.0044)\\
\hspace*{6mm}$\gamma$ & 0.0325 & (0.0059) & 0.0277 & (0.0025) & 0.0317 & (0.0038) & 0.0347 & (0.0041)\\
\hspace*{6mm}$\lambda$ & -0.0257 & (0.0056) & -0.0261 & (0.0038) & -0.0348 & (0.0066) & -0.0410 & (0.0068)\\
\addlinespace
\multicolumn{9}{l}{\hspace*{0mm}Production function and price discount}\\
\addlinespace
\hspace*{6mm}$\delta$ & 0.3756 & (0.0249) & 0.3756 & (0.0412) & 0.3756 & (0.0396) & 0.3756 & (0.0333)\\
\hspace*{6mm}$A$ & 4.1435 & (0.0854) & 4.1064 & (0.1318) & 4.1040 & (0.0976) & 4.1036 & (0.0766)\\
\hspace*{6mm}$\alpha_{H_0}$ & 0.0220 & (0.0088) & 0.0298 & (0.0177) & 0.0337 & (0.0195) & 0.0344 & (0.0118)\\
\hspace*{6mm}$\alpha_{male}$ & 0.0086 & (0.0016) & 0.0074 & (0.0016) & 0.0074 & (0.0016) & 0.0074 & (0.0016)\\
\hspace*{6mm}$\beta$ & 0.0725 & (0.0158) & 0.0752 & (0.0169) & 0.0753 & (0.0162) & 0.0753 & (0.0138)\\
\hspace*{6mm}$\sigma_{\epsilon}$ & 0.0097 & (0.0010) & 0.0100 & (0.0014) & 0.0100 & (0.0014) & 0.0100 & (0.0014)\\
\addlinespace
\multicolumn{9}{l}{\hspace*{0mm}Measurement error}\\
\addlinespace
\hspace*{6mm}$\sigma_{\eta}$ & 0.3823 & (0.0151) & 0.3827 & (0.0152) & 0.3830 & (0.0153) & 0.3830 & (0.0152)\\
\hspace*{6mm}$\sigma_{\iota}$ & 0.0427 & (0.0017) & 0.0425 & (0.0017) & 0.0425 & (0.0017) & 0.0425 & (0.0017)\\
\addlinespace
\bottomrule
\addlinespace[0.5em]
\multicolumn{9}{p{1\textwidth}}{\parbox[t]{1\textwidth}{\footnotesize{\emph{Note}: Estimation results from re-estimating the model under different reference points standard deviation assumptions. See Section \ref{sec:varyrefesti} for discussions.}}}\\
\addlinespace
\end{tabular}
\end{adjustbox}
\end{table}

\clearpage
\pagebreak

\appendix

\setlength{\footnotemargin}{5.75mm}
\begingroup
\doublespacing
\centering
\Large ONLINE APPENDIX \\
\Large\begin{singlespace}\href{\PAPERDOIURL}{\PAPERTITLE}\end{singlespace}
\large \AUTHORWANG{},
\AUTHORPUENTES{},
\AUTHORBEHRMAN{},
\AUTHORCUNHA\\[1.0em]
\endgroup



\renewcommand{\thefigure}{A.\arabic{figure}}
\setcounter{figure}{0}
\renewcommand{\thetable}{A.\arabic{table}}
\setcounter{table}{0}
\renewcommand{\theequation}{A.\arabic{equation}}
\setcounter{equation}{0}
\renewcommand{\thefootnote}{A.\arabic{footnote}}
\setcounter{footnote}{0}

\section{Solution and Estimation Details (Online Publication)}

\subsection{Reference Point and Preference for Height \label{sec:apprefsd}}

\subsubsection{The Optimal Nutritional-Choice Problem\label{sec:fullproblem}} 
Given Equation \eqref{eq:budget_constraint} for the budget constraint, Equation \eqref{eq:prod_function} for the production function, Equation \eqref{eq:preferences} for preferences, and the reference points distributions discussed in Section \ref{sec:information}, the nutritional-choice problem is:
\begin{align}
    \begin{gathered}\label{eq:optimization}
        \max_{N} 
        \left\{
            C + \rho\cdot C^{2} + 
            \gamma \cdot H^{24} + 
            \lambda \cdot \int_{R_{yv}} \left(H^{24}-R_{yv}\right)\cdot\mathbbm{1}\left\{ H^{24}> R_{yv}\right\}dF(R_{yv})
        \right\}
        \\
        C + p^{N}_{yv}\cdot(1-\delta\cdot\mathbbm{1}\left\{v=Atole\right\})\cdot N = Y\\
        H^{24} = \exp(A+ X\cdot\alpha+\epsilon)\cdot N^{\beta}
    \end{gathered}
\end{align}
For notational convenience, we drop subscripts, let $\hat{A}=\exp\left(A + X\cdot\alpha + \epsilon\right)$, replace $H^{24}$ and $C$ as functions of $N$, and rewrite Equation \eqref{eq:optimization} as:
\begin{align}
    \begin{gathered}\label{eq:optimization:single}
        \max_{N} 
        \left\{
        \begin{array}{c}
            \left(Y - p \cdot N\right) + 
            \rho\cdot\left(Y - p \cdot N\right)^{2} \\
            + \gamma \cdot \hat{A} \cdot N^{\beta} \\
            + \lambda \cdot \left(\hat{A} \cdot N^{\beta}-\mu_R\right)  \cdot
                    \Phi\left(\frac{\hat{A} \cdot N^{\beta}-\mu_R}{\sigma_R}
                \right) \\
            + \lambda \cdot \sigma_R \cdot \phi\left(
                \frac{\hat{A} \cdot N^{\beta}-\mu_R}{\sigma_R}
                \right) \\
        \end{array}
        \right\}
    \end{gathered}
\end{align}
In Equation \eqref{eq:optimization:single}, we analytically integrate the integral from Equation \eqref{eq:optimization}.\footnote{The expected utility function contained the expectation of a truncated normal random variable:
\begin{multline}
	\int_{R_{yv}}
	\left(H^{24}-R_{yv}\right)\cdot
	\mathbbm{1}\left\{ H^{24}> R_{yv}\right\}dF(R_{yv})=  \left(H^{24}-\mu_{R_{yv}}\right)\cdot\left(\Phi\left(\frac{H^{24}-\mu_{R_{yv}}}{\sigma_{R_{yv}}}\right)\right)
	+\sigma_{R_{yv}}\cdot\phi\left(\frac{H^{24}-\mu_{R_{yv}}}{\sigma_{R_{yv}}}\right) \label{eq:truncatedNormal}
\end{multline}}

\subsubsection{Marginal Benefits and Marginal Costs\label{sec:foc}} 
Solving Equation \eqref{eq:optimization:single}, from first-order conditions, we have the following optimality condition:
\begin{align}
    \begin{gathered}\label{eq:foc}
        \underbrace{
        p + 2 \cdot \rho \cdot p \cdot Y
        - 2 \cdot \rho \cdot p^2 \cdot N}_{
        \substack{
             \text{Marginal utility costs of nutrition} \\
        }}
        =
        \overbrace{
        \underbrace{
        \beta \cdot \hat{A} \cdot N^{\beta-1}}_{
        \substack{
             \text{Marginal product of} \\
             \text{nutrition on height}
        }}
        \cdot 
        \underbrace{
        \left(
            \gamma + 
            \underbrace{\lambda \cdot 
            \Phi\left(
                \frac{\hat{A} \cdot N^{\beta}-\mu_R}{\sigma_R}
                \right)}_{
                    \substack{
                     \text{Reference points effects} \\
                }}
        \right)}_{
        \substack{
             \text{Marginal expected utility} \\
             \text{of height}\\
             \text{given reference points}\\
        }}}^
        {\text{Marginal expected utility benefits of nutrition}}
        \thinspace\thinspace,
        \\
    \end{gathered}
\end{align}
which is obtained after canceling out equivalent terms.\footnote{For details on derivation, see \href{https://fanwangecon.github.io/Py4Econ/math/derivative/htmlpdfr/fs_diff_normcdf.html}{here}. This is the interior optimality condition. Given quadratic utility on $C$ as well as potential marginal losses beyond $\mu_R$, households might choose corners. Given our estimated parameters, in particular $\rho$, and the joint income and price distribution for households, empirical optimal choices are in the interior.}$^{,}$\footnote{Equation \eqref{eq:foc} does not permit an analytical expression for the optimal nutritional choices $N$.} 

Equation \eqref{eq:foc} equates the marginal utility costs and benefits of additional nutritional inputs $N$.
If $\lambda = 0$, the term associated with reference points disappears. Given $\gamma>0$, if $-\gamma < \lambda < 0$, marginal benefits are positive and decreasing in $N$.\footnote{Since $0 < \beta <1$, the marginal product of nutrition on height is decreasing in $N$. Additionally, with $\lambda < 0$, the marginal expected utility of height is decreasing in $N$: 
$$
\frac{\partial\left(\gamma + \lambda\cdot \Phi\left(\frac{\hat{A}\cdot N^{\beta}-\mu_R}{\sigma_R}\right)\right)}{\partial N} = \lambda\cdot\left(\frac{\beta \cdot \hat{A} \cdot N^{\beta-1}}{\sigma_R}\right)\cdot\phi\left(\frac{\left(\hat{A}\cdot N^{\beta}-\mu_R\right)}{\sigma_R}\right) < 0
\thinspace\thinspace.
$$
With $-\gamma < \lambda < 0$, as $N$ increases, marginal expected utility of height remains positive:
$$\lim_{N\rightarrow \infty} \left(\gamma + \lambda \cdot \Phi \left(\frac{\hat{A}\cdot N^{\beta} - \mu_R}{\sigma_R}\right)\right) = \gamma + \lambda > 0
\thinspace\thinspace.
$$} If $\lambda < -\gamma < 0$, as expected height increases beyond $\mu_R$, marginal benefits become marginal losses. 

Consider the reference points component of marginal benefits of $N$. 
First, marginal benefits are increasing in $\lambda$. 
Second, with $\lambda<0$, the marginal benefits are increasing in $\mu_R$ and decreasing in $N$: marginal benefits diminish as expected height approaches $\mu_R$. 
Third, as uncertainty increases, preferences become more linear in height: $$
\lim_{\sigma_R \rightarrow \infty} \left( \gamma + \lambda \cdot \Phi\left( \frac{\hat{A}\cdot N^{\beta} - \mu_R}{\sigma_R} \right)\right) = \gamma + \lambda \cdot 0.5 
\thinspace\thinspace.
$$
In effect, as $\sigma_R$ increases towards infinity, predictions from models with and without reference points can not be distinguished. Fourth, the derivative of the normal CDF with respect to its $\sigma_R$ has alternating signs depending on the relative values of $\hat{A}\cdot N^{\beta}$ and $\mu_R$:
$$
\frac{\partial\left(\gamma + \lambda\cdot \Phi\left(\frac{\hat{A}\cdot N^{\beta}-\mu_R}{\sigma_R}\right)\right)}{\partial\sigma_R} = (-1)\cdot\lambda\cdot\left(\frac{\hat{A}\cdot N^{\beta}-\mu_R}{\sigma_R^2}\right)\cdot\phi\left(\frac{\hat{A}\cdot N^{\beta}-\mu_R}{\sigma_R}\right)
\thinspace\thinspace.
$$ It turns out that increasing $\sigma_R$ can increase or decrease optimal choices depending on the relative values of $\gamma$ and $\lambda$. We illustrate the effects of changing $\sigma_R$ on optimal choices in Figure \ref{fig:refpointhgtutil}, which we discuss in the next set of paragraphs.

\subsubsection{Optimal Nutritional Choices and $\sigma_R$\label{sec:sigmaopti}}

Panels a, c, and e of Figure \ref{fig:refpointhgtutil} show health (height) at age 24 months along the x-axis and household consumption along the y-axis. We plot the shape of indifference curves for a given value of $\gamma$ and different values of $\lambda$ and $\sigma_R$ together with the consumption-health possibility frontier, which combines the budget constraint with the production function for height.  

Panels b, d, and f show the difference between height at age 24 months and the mean beliefs about the reference height along the x-axis, and the component of utility that  includes the reference point for height along the y-axis. We believe it is helpful to zero in on this component because it is the primary driver of our empirical analysis, and it shows the role that the parameters $\lambda$ and $\sigma_R$ play in our study of the determinants of height at age 24 months.  

In panels a and b, we use our estimated values of $\gamma$ and $\lambda$ for the scenario in which $\sigma_R = 0.5$. In panels c and d, we set $\lambda$ to be negative $1.5$ times the estimated value of $\gamma$. In panels e and f, the $\lambda$ value is lowered to be negative $2.5$ times the estimated value of $\gamma$. 

For the configuration of values of $\gamma$ and $\lambda$ in panel a, an increase in $\sigma_R$ increases optimal height because the marginal benefits of height levels that exceed $\mu_{R}$ remains positive as panel b shows. However, for the configuration of values of $\gamma$ and $\lambda$ in panel c, we have a bliss point for height that varies just a little as we increase $\sigma_R$. In panel e, as $\lambda$ continues to decrease, the disutility of height beyond the mean reference point decreases so fast that the uncertainty becomes costly. As a result, an exogenous increase in $\sigma_R$ reduces the optimal level of height. 

\newcommand{\subFigWidthApp}{0.50}
\begin{figure}[hbt!]
    \centering
    \begin{subfigure}[b]{\subFigWidthApp\textwidth}
        \centering
        \includegraphics[width=\linewidth]{\string"fig_main/fig1d_Tester_Indiff_person_495_refsd_increasem3_20210130".eps}
        \caption{Choices with $\gamma=0.0325$, $\lambda=-0.0257$}
    \end{subfigure}~
	\begin{subfigure}[b]{\subFigWidthApp\textwidth}
	    \centering
    	\includegraphics[width=\linewidth]{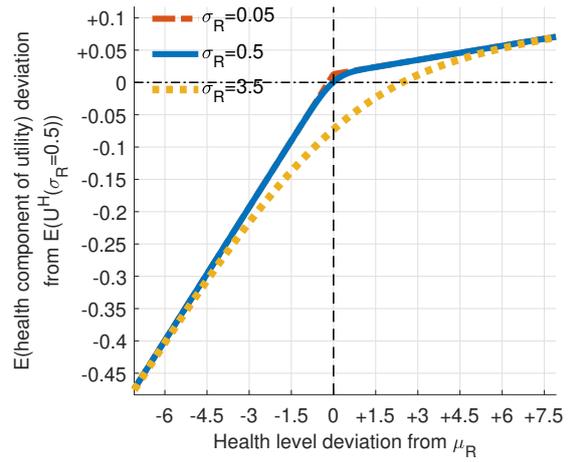}
    	\caption{Health preference with $\gamma=0.0325$, $\lambda=-0.0257$}
	\end{subfigure}
	\par\medskip
	\begin{subfigure}[b]{\subFigWidthApp\textwidth}
        \centering
        \includegraphics[width=\linewidth]{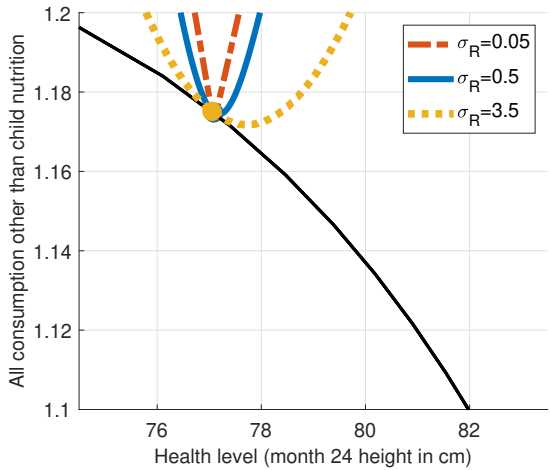}
        \caption{Choices with $\gamma=0.0325$, $\lambda=-\gamma\cdot1.5$}
    \end{subfigure}~
	\begin{subfigure}[b]{\subFigWidthApp\textwidth}
	    \centering
	    \includegraphics[width=\linewidth]{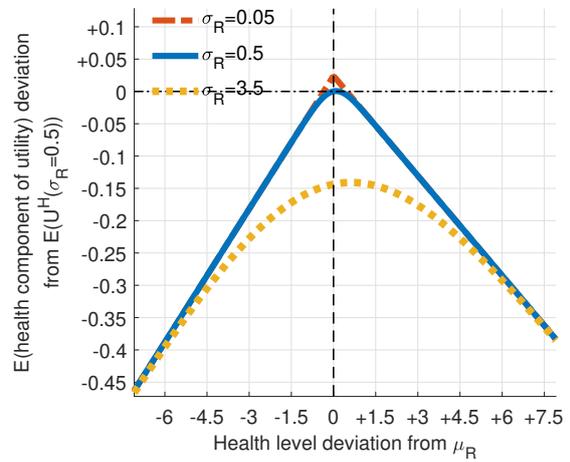}
	    \caption{Health preference with $\gamma=0.0325$, $\lambda=-\gamma\cdot1.5$}
	\end{subfigure}
	\par\medskip
	\begin{subfigure}[b]{\subFigWidthApp\textwidth}
        \centering
        \includegraphics[width=\linewidth]{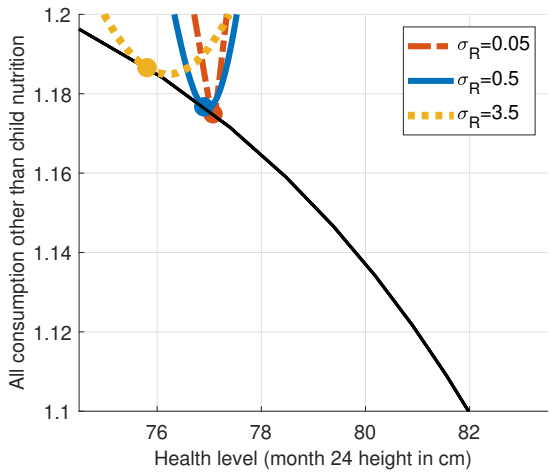}
        \caption{Choices with $\gamma=0.0325$, $\lambda=-\gamma\cdot2.5$}
    \end{subfigure}~
	\begin{subfigure}[b]{\subFigWidthApp\textwidth}
	    \centering
	    \includegraphics[width=\linewidth]{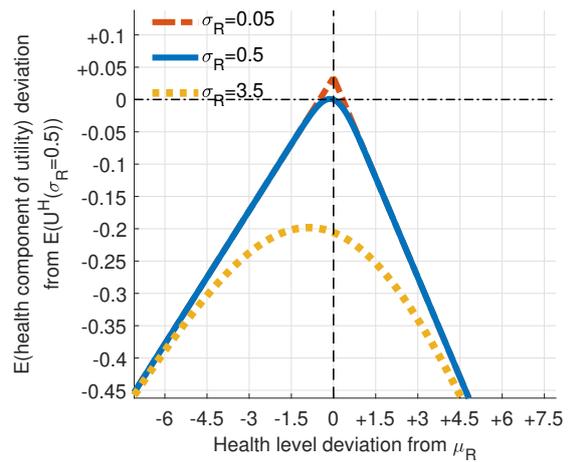}
	    \caption{Health preference with $\gamma=0.0325$, $\lambda=-\gamma\cdot2.5$}
	\end{subfigure}	
	\caption{Reference points distribution standard deviation ($\sigma_R$) and health outcomes}
	\label{fig:refpointhgtutil}
\end{figure}
\pagebreak

\clearpage
\pagebreak
\subsection{Model Solution and Estimation \label{sec:solu}}

\subsubsection{Model Solution\label{sec:modelsolu}}
We use an iterative grid search routine that integrates over the two shocks facing the household ($R_{yv}$ and $\epsilon$) to solve the model. First, for each household, given $\Omega_{yv}$, we construct a grid with $Q_{1}$ points for protein intake. The points range from zero protein intake up to the maximum that each household could purchase.

Second, we assume that $R_{yv}$ and $\epsilon$ follow independent normal distributions. We draw $M$ productivity shocks $\epsilon$ for each household from $ \mathcal{N}(0,\, \sigma_{\epsilon}^{2})$. 

Third, for each household and each $\epsilon$ shock drawn, we compute the integral over the distribution of $R_{yv}$ analytically because we can use the fact that the component of the utility function that depends on reference point involves the expectation of a truncated normal random variable as shown in Equation \eqref{eq:optimization:single}. Thus, each point on the choice grid has an expected utility value associated with it. Then, we find the grid point that has the maximum expected utility value. 

Fourth, we construct a new finer household-specific choice grid with $Q_{2}$ points around the optimal nutritional choice from the initial $Q_{1}$ point grid. We repeat steps one through four to evaluate the expected utility function and find the maximum as before. The process is iterated for $Z$ iterations until the total difference in optimal nutritional choices between iterations meets the convergence criteria. This solution provides the exact optimal choices for each household. Given our estimation problem, the speed of obtaining the likelihood function given each set of parameters is determined by $M$, the number of productivity shocks that we draw. For $M$ less than 50, we can evaluate the likelihood within seconds.

\subsubsection{Likelihood}

Let $\ln \left(H_{i}^{24,*}\right)$ and $\ln\left( H^{24}(\Omega_{yvi}; \theta)\right)$ denote the log of observed and model predicted height at age 24 months, respectively. Similarly, let $\ln \left(N^{*}_{i}\right)$ and $\ln\left( N(\Omega_{yvi}; \theta)\right)$ denote the log of observed and model predicted protein intake. We note that the model predicted values depend on the vector of parameters $\theta$: 

\begin{equation*}
    \theta=\{\underbrace{\rho, \gamma, \lambda}_{\text{Preference}}, \overbrace{\delta}^{\substack{\text{Atole} \\ \text{disc.}}}, \underbrace{A, \alpha , \beta, \sigma_{\epsilon} }_{\substack{\text{Production} \\ \text{Function}}}, \sigma_{\eta},\sigma_{\iota}\}
\end{equation*}

According to our discussion in Section \ref{sec:estiidentify}, the likelihood is based on the model optimal and observed nutritional choices, as well as the model simulated and observed heights at 24 months of age:
\begin{equation}
\label{eq:likelihood}
\max_{\theta\in\Theta} \sum_{y=1970}^{1975}\sum_{v} \bigg\{ \sum_{i=1}^{n_{yv}} \log \bigg( \int_{\epsilon} \phi_{\iota}
\big(\ln H_{24,i}^{*}-\ln H^{24}(\Omega_{yvi}; \theta)\big)
\cdot \phi_{\eta}\big(\ln N^{*}_{i} - \ln N(\Omega_{yvi}; \theta)\big) dF(\epsilon_i) \bigg)\bigg\}
\end{equation}

To find the $\theta$ that maximizes the likelihood, we search across parameter space using quasi-Newton methods, and initiate the likelihood with a range of parameter values to find the global maximum.\footnote{Initial values of the parameters: for preference parameters, we investigate $\rho$ equals to or less than $0$, and we test a range of $\gamma$ values, with corresponding $\lambda$ values that make preference in height linear, or have different degrees of concavity. For the Atole discount $\delta$ parameter, we test from 10 to 90 percent discounts at 10 percent intervals. We start production function parameters at the same values always, which are parameters estimated from a instrumental variable regression in which the Atole dummy is an instrument for protein intakes.} We obtain standard errors from the approximated inverse Hessian.

\subsection{Alternative Assumptions about Variance Beliefs} \label{sec:appsecalter}

Our model assumes that parents adopt the expected height of two-year-old children in their village as the reference point. Parents estimate the expected height using the observations of the children in their village. We assume that their estimation of the expected height follows a normal distribution with mean belief parameter $\mu_{R_{y,v}}$ and variance belief parameter $\sigma_{R{y,v}}^{2}$. In our empirical analysis, we assume that $\mu_{R_{y,v}}$  captures the uncertainty about the mean belief, so  $\sigma_{R{y,v}} = 0.5$.

In our sensitivity analysis, we adopt the assumption that the variance belief is equal to the variance of height. We argue that this approach is an upper bound for variance beliefs. Under this assumption, $\sigma_{R{y,v}} = 3.5$. To complete our sensitivity analysis, we also investigate our model's decomposition findings for scenarios in which $\sigma_{R{y,v}} = 1.5$ and $\sigma_{R{y,v}} = 2.5$. 

\subsection{Data for Estimation \label{sec:estidata}}

We include in the estimation children who were born between 1970 and 1975. We show summary statistics for these children in Sections \ref{sec:stat} and \ref{sec:gap}. As discussed earlier, we do not observe both initial heights and heights at month 24 for children born before or after these years.

We use the months 15 to 24 average protein intakes, heights at month 24, protein prices, incomes, gender, and initial height variables shown in Table \ref{summcovarmain} and described in Section \ref{sec:stat} as $N$, $H^{24}$, $p_{yv}^{N}$, $Y$, and components of $X$. We use protein intakes because \textcite{puentes_early_2016} shows that proteins rather than non-protein components of calories matter for height growth in these INCAP data. Ideally, our intake variable should be averaged from month 0 to month 24. However, we do not observe protein values from month 0 to month 12 for close to half of our sample because data collection for this age group started in 1973. Also, it is not easy to estimate the protein component of breastmilk for children who rely on breastfeeding in the first year of life.

In terms of reference points, controlling for gender, we use the predicted value of the linear trends across cohorts between 1970 to 1975 -- the trends and coefficients are shown in Figure \ref{fig:PortHgtGap} and described in Section \ref{sec:gap} -- as the reference points for Atole and Fresco villages. Specifically, we use the linear trends from panels a and b of Figure \ref{fig:PortHgtGap} along with a gender adjustment. The trend for Atole villages could also be approximated with a quadratic trend, but switching to quadratic trends has minimal effects on estimated parameters. Potentially, we could also use the local polynomial approximated nonlinear trends as reference points, but the linear trends as shown in Figure \ref{fig:PortHgtGap} closely approximate the local polynomial trends, which further could be fluctuating due to sample variation for each birth cohort group. This provides us with a set of village, cohort-calendar-year and gender-specific predictions of height at 24 months of age: $E(H^{24}|\mathrm{year},\mathrm{gender},\mathrm{atole})=\phi_0+\phi_1\cdot\mathrm{year}+\phi_2\cdot\mathrm{atole}\cdot\mathrm{year}+\phi_3\cdot\mathrm{gender}$. We use the 24 months of age height predictions to obtain $\mu_{R_{\mathrm{year},\mathrm{gender},\mathrm{atole}}}$. For example, the predicted linear-trend value that corresponds to the height at month 24 of those born in 1970 is the mean reference point for the cohort born in 1972: $\mu_{R_{1972,\mathrm{gender},v}}=E(H^{24}|1970,\mathrm{gender},v)$. By construction, if simulated results from our estimated model match the average 24 months of age heights for different cohorts, we will also have matched the mean reference point values.

For the main results presented in the paper, we fix $\sigma_{R_{\mathrm{year},\mathrm{gender},\mathrm{atole}}} =0.5$ for Atole and Fresco villages in all years. We do this because the standard error of mean height across cohorts and villages is generally around 0.5~cm. The average standard error of heights in Fresco villages is 0.60, 0.59, and 0.51~cm for the 1970-71, 1972-73, and 1974-75 cohorts, respectively. The average standard errors of height in Atole villages are 0.67, 0.45, and 0.48~cm for the 1970-71, 1972-73, and 1974-75 cohorts, respectively.

For our alternative assumption for the standard deviation of the reference points discussed in Appendix Section \ref{sec:appsecalter}, we fix $\sigma_{R_{\mathrm{year},\mathrm{gender},\mathrm{atole}}} =3.5$ for Atole and Fresco villages in all years. We do this because height variances across cohorts and villages do not seem to vary systematically. The standard deviations of heights in Fresco villages are 3.33, 3.34, and 3.24~cm for the 1970-71, 1972-73, and 1974-75 cohorts, respectively. The standard deviation of height in Atole villages are 3.13, 3.73, and 3.53~cm for the 1970-71, 1972-73, and 1974-75 cohorts, respectively.

\subsection{Additional Estimation Information \label{sec:estimorer}}

\subsubsection{Additional Model fit}
\begin{table}[htbp]
\renewcommand{\arraystretch}{1.3}
\centering
{\small
\def\sym#1{\ifmmode^{#1}\else\(^{#1}\)\fi}
\caption{Model Predictions from estimating the model under different assumptions about the standard deviation of reference points distribution $\sigma_{R}$.\label{tab:paramfitmulti}}
\begin{adjustbox}{max width=1\textwidth}
\begin{tabular}{lcccc|cccc}
\toprule
& \multicolumn{4}{C{6cm}}{\hspace*{-10mm}\textbf{Average protein choices}} & \multicolumn{4}{C{6cm}}{\hspace*{-15mm}\textbf{Average height outcome}} \\
\cmidrule(l{5pt}r{5pt}){2-5} \cmidrule(l{5pt}r{5pt}){6-9}
& \multicolumn{1}{C{1.5cm}}{\textit{\small$\sigma_R=0.5$}} & \multicolumn{1}{C{1.5cm}}{\textit{\small$\sigma_R=1.5$}} & \multicolumn{1}{C{1.5cm}}{\textit{\small$\sigma_R=2.5$}} & \multicolumn{1}{C{1.5cm}}{\textit{\small$\sigma_R=3.5$}} & \multicolumn{1}{C{1.5cm}}{\textit{\small$\sigma_R=0.5$}} & \multicolumn{1}{C{1.5cm}}{\textit{\small$\sigma_R=1.5$}} & \multicolumn{1}{C{1.5cm}}{\textit{\small$\sigma_R=2.5$}} & \multicolumn{1}{C{1.5cm}}{\textit{\small$\sigma_R=3.5$}} \\      
\midrule
\multicolumn{9}{l}{\hspace*{0mm}\textbf{\normalsize Panel a: Average}} \\
Fresco          &  19.38&   19.24&   19.25&   19.29&   76.77&   76.77&   76.78&   76.77\\
Atole           &  25.10&   25.30&   25.30&   25.47&   78.19&   78.32&   78.33&   78.39\\
\midrule
\multicolumn{9}{l}{\hspace*{0mm}\textbf{\normalsize Panel b: Averages across genders}} \\
Fresco Female   &  18.15&   18.14&   18.16&   18.15&   76.07&   76.11&   76.12&   76.11\\
Fresco Male     &  23.96&   24.17&   24.17&   24.24&   77.60&   77.74&   77.75&   77.81\\
Atole Female    &  20.48&   20.23&   20.23&   20.30&   77.40&   77.36&   77.36&   77.35\\
Atole Male      &  26.16&   26.34&   26.33&   26.60&   78.74&   78.85&   78.86&   78.93\\
\midrule
\multicolumn{9}{l}{\hspace*{0mm}\textbf{\normalsize Panel c: Averages across years}} \\
Fresco 1970-71  &  18.43&   18.61&   18.63&   18.61&   76.52&   76.59&   76.60&   76.57\\
Atole  1970-71  &  23.13&   23.52&   23.46&   23.46&   77.71&   77.85&   77.85&   77.88\\
Fresco 1972-73  &  19.30&   19.22&   19.24&   19.25&   76.80&   76.79&   76.79&   76.79\\
Atole  1972-73  &  24.51&   24.76&   24.76&   24.92&   78.07&   78.20&   78.20&   78.27\\
Fresco 1974-75  &  19.97&   19.61&   19.61&   19.69&   76.89&   76.86&   76.86&   76.85\\
Atole  1974-75  &  27.13&   27.13&   27.15&   27.46&   78.65&   78.78&   78.80&   78.87\\
\bottomrule
\addlinespace[0.5em]
\multicolumn{9}{p{1\textwidth}}{\parbox[t]{1\textwidth}{}}\\
\end{tabular}\end{adjustbox}
}
\end{table}

In Table \ref{tab:paramfitmulti}, we provide additional information on the fits of the estimated model under varying assumptions about $\sigma_{R_{y,v}}$. Model fits with respect to protein choices and height outcomes are presented across the columns, and model fits with respect to overall averages in Atole and Fresco villages, as well as average by gender and time are presented in separate rows. 

Estimated models across $\sigma_{R_{y,v}}$ assumptions -- with estimates shown in Table \ref{tab:paramesti} -- are all able to provide similarly good fits between model predictions and key data moments.

\subsubsection{Additional Estimates}

\begin{figure}[htbp]
\makebox[\textwidth][c]{\includegraphics[width=1.10\textwidth]{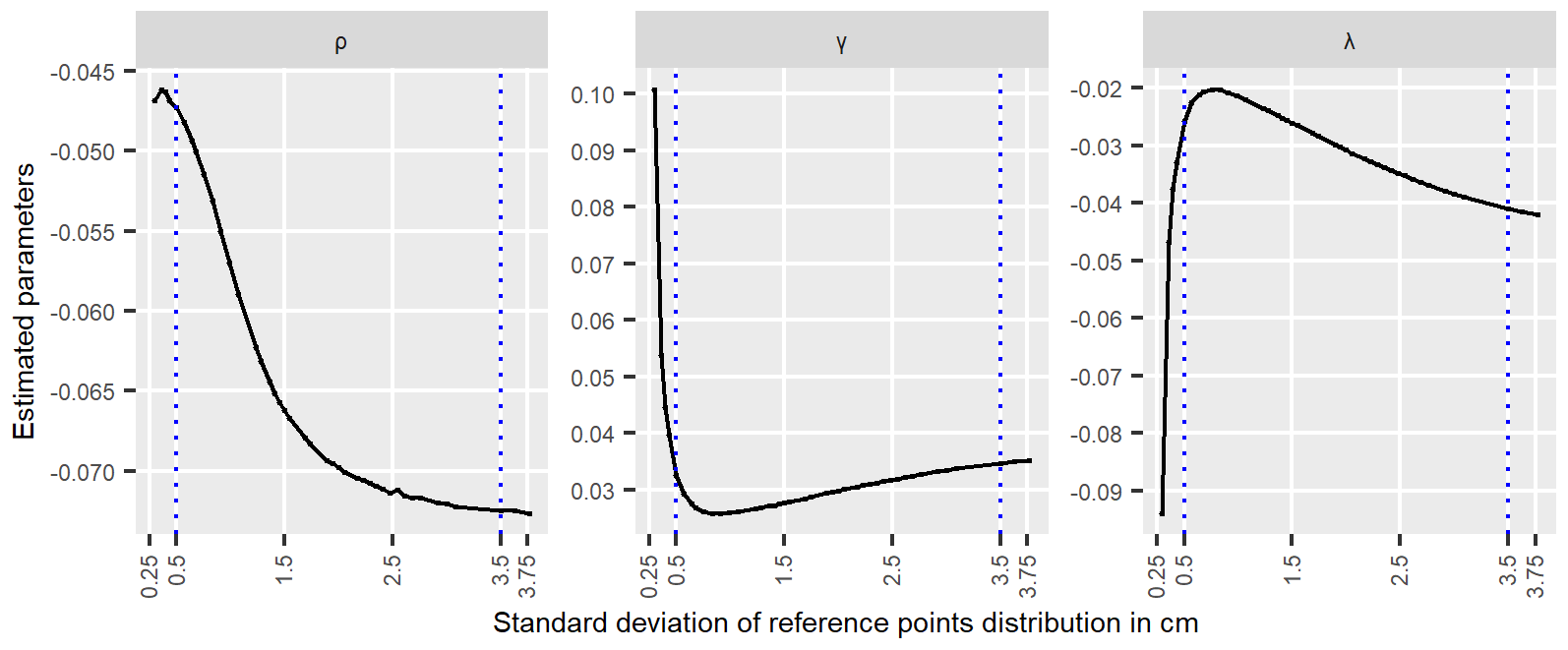}}
\caption{Preference parameter estimates from estimating the model under different assumptions about the standard deviation of reference points distribution $\sigma_{R_{y,v}}$}
\label{fig:estimatesmulti}
\end{figure}

In Table \ref{tab:paramesti}, we present parameter estimates under four assumptions for $\sigma_{R_{y,v}}$ from 0.5~cm to 3.5~cm. In this section, for additional information, we provide key estimates from estimating the model at $\sigma_{R_{y,v}}$ values from 0.25~cm to 3.75~cm at 0.05~cm intervals. In Figure \ref{fig:estimatesmulti}, we visualize the point-estimates results for the core preference parameters of $\rho$, $\gamma$, and $\lambda$ in three sub-figures. We mark with dashed blue lines values of $\sigma_{R_{y,v}}$ along the x-axis where $\sigma_{R_{y,v}}=0.5$ and $\sigma_{R_{y,v}}=3.5$. 

For $\rho$, the quadratic coefficient on non-nutritional consumptions, the estimated coefficient becomes more negative as $\sigma_{R_{y,v}}$ increases. This finding implies lower marginal gains from additional non-nutritional consumption at higher $\sigma_{R_{y,v}}$.

$\gamma$ and $\lambda$ change in a symmetric fashion. Between $\sigma_{R_{y,v}}=0.5$ and $\sigma_{R_{y,v}}=3.5$, $\gamma$ increases with $\sigma_{R_{y,v}}$, and $\lambda$ decreases with $\sigma_{R_{y,v}}$. These correspond to higher marginal gains from additional height, before the mean of the reference points distribution, at higher $\sigma_{R_{y,v}}$. Moreover, these also correspond to a sharper decline in marginal gains from additional height, after the mean of the reference points distribution, at higher $\sigma_{R_{y,v}}$. As noted in our discussion of Table \ref{tab:paramesti}, preferences past the mean reference point are slightly increasing ($\gamma > - \lambda$) at lower $\sigma_{R}$ values, and slightly decreasing ($\gamma < - \lambda$) at higher $\sigma_{R_{y,v}}$ values.

The changes in parameter estimates as we vary $\sigma_{R_{y,v}}$ reflect an effort by the estimation routine to fit the data. As $\sigma_{R_{y,v}}$ increases from 0.5~cm to 3.5~cm, increases in $\gamma$ estimates and decreases in $\rho$ estimates increase incentives to invest in child health, but reductions in $\lambda$ estimates reduce the incentives to invest in child health above the reference point. Overall, the estimated models across $\sigma_{R_{y,v}}$ assumptions provide a good fit with the data as shown in Table \ref{tab:paramfitmulti}.

We note that below $\sigma_{R_{y,v}} <0.25$, $\gamma$ parameter estimates increase sharply and $\lambda$ parameter estimates drop sharply. While model fits from the estimated parameters between $\sigma_{R_{y,v}}=0.5$ and $\sigma_{R_{y,v}}=3.5$ are similar, the fit is significantly worse when $\sigma_{R_{y,v}}$ falls below 0.4~cm. 

\clearpage
\pagebreak


\section{Additional Counterfactual Exercises (Online Publication)}

\renewcommand{\thefigure}{B.\arabic{figure}}
\setcounter{figure}{0}
\renewcommand{\thetable}{B.\arabic{table}}
\setcounter{table}{0}
\renewcommand{\theequation}{B.\arabic{equation}}
\setcounter{equation}{0}
\renewcommand{\thefootnote}{B.\arabic{footnote}}
\setcounter{footnote}{0}

\subsection{Targeted versus Universal Policy Experiments\label{sec:targetuniversal}}

We evaluate the impacts of counterfactual policy experiments that target protein-price discounts towards the most impoverished children and a universal policy that provides standard price discounts for all, given a fixed budget. The analysis in this section allows us to exploit dynamic features of our model with reference points to distinguish among three possible effects of protein-price-subsidy policies: 1, the direct impact of subsidies on treated children; 2, the indirect effect of subsidies via shifting reference points on treated children; 3, the indirect impact of shifting reference points on untreated children.

Our outcome of interest remains height at 24 months of age. There has been a long debate in both the development and early childhood literature about the trade-offs between targeted vs. universal subsidy policies \autocite{besley_principles_1990, gelbach_more_1997, coady_targeting_2004}. Under universal subsidy policies, the entire population benefits. Under targeted policies, subsidies are generally given to those deemed (means-tested for) the most in need.

Within our empirical setting, our first key result here is that the universal policy -- which is generally considered to be insufficiently beneficial for the poor -- is much better for the most impoverished children in later than earlier cohorts because their heights increase substantially in later cohorts due to aggregate increases in reference points. Given our estimates, the gaps between targeted and universal policies on the poor are smaller once we consider the endogenous evolution of reference points. 

Our second key result is that, after an initial period in which only targeted individuals benefit, the targeted policy -- which would typically only benefit the poor -- has substantial positive effects on the non-targeted most affluent children through the externality of reference point changes.

Both results show that reference points amplify targeted and universal subsidy policies on all children within the same village. We analyze the height distributions induced by budget-equivalent policies that gradually increase the proportion of children receiving price subsidies. The budget is equal to the cost of providing the observed levels of consumption of health-center-provided free proteins in the Atole villages. Overall, we find that within our empirical context, all policies induce similar mean heights. However, the policies generate substantial differences in height variances: the most-targeted and the most-universal policies both increase variances. Within the set of budget-balancing policies we consider, we find that variance in height is minimized when 70 percent of the most-impoverished children receive 18 percent price discounts.

In Section \ref{sec:varyrefesti}, we show model estimates under varying assumptions on the standard deviation of the reference points distribution. Similar to the decompositional analysis, targeted and universal policies are similar when we use the estimates from different columns of Table \ref{tab:paramesti}. 

\subsubsection{Simulation Design -- Budget Balancing Targeted to Universal Policies \label{sec:budgetbalance}}

In Section \ref{sec:targetuniversal}, we compare policies in which increasingly more significant fractions of individuals in a village are targeted as subsidy recipients. To keep subsidy costs the same across policies, we reduce the subsidies provided to targeted children as the fraction of children targeted increases. We conduct policy counterfactuals that target children by annual household income.

To simulate the policies, we draw 500 individuals based on the empirical joint distribution of incomes, gender, and initial heights of all children from Atole villages. Then, we start at the reference point from Atole villages in the year 1970 and simulate our model forward. Each model period is two years, and we simulate the model four times to obtain results for the 1970, 1972, 1974, and 1976 cohorts. The state-space distribution and protein prices facing households are the same across cohorts, but each cohort faces different endogenously evolving reference points\footnote{We shift the mean of the reference points distribution. Under the assumption in Appendix Section \ref{sec:appsecalter} the standard deviation of the reference points is based on the variance of the height distribution, this variance, as discussed in Appendix Section \ref{sec:estidata}, is stable across cohorts. Under the assumption in Section \ref{sec:evolution}, since the number of new-borne is similar across cohorts, the standard error of the mean estimate of reference points distribution would also be stable across cohorts.}, which lead to variations in nutritional choices and heights.\footnote{The goal is to isolate the effects of price subsidy and reference point changes, and abstract away from other potential observed differences in initial heights, non-protein prices, and incomes. If the price discount policy were 38 percent and provided to all individuals, the height path would be similar to the observed height path, but it would not be identical because all of our simulated cohorts are identical except for their reference points.}

Following the model interpretation of the protein supplementation policy as a price discount, our counterfactuals here involve changing that discount.\footnote{Alternatively, we could provide households with different levels of protein transfers, but that would involve forcing households to consume a fixed level of subsidized proteins.} Under the universal policies, all families receive subsidies through a standard price discount. Under the policies that target the poor, we transfer the subsidies that rich children received under the universal policy to poor children by increasing the price discount that the poor receive. Specifically, let $\tau$ be the fraction of most impoverished children receiving price discounts and $\delta$ be the percentage price discount that children receive. $Z\left(\tau,\delta\right)$ is the total cost of a subsidy in grams of protein for 1970, 1972, 1974, and 1976 cohorts, given height distribution $\Gamma$ for each cohort:
\begin{equation}
\label{eq:targetcost}
Z\left(\tau,\delta\right) =
\sum\limits_{
	\substack{
	\mathrm{cohort} \\ \in{\left\{70,72,74,76\right\}}}
	}
\left\{\delta\cdot
\int_{\epsilon}
\int_{Y_{min}}^{F_{Y}^{-1}\left(\tau\right)}
\int_{X}
N\Big(
\substack{
	Y,X,\epsilon; \\
	\delta, \Gamma_{\mathrm{cohort}}
}
\Big)f\left(X|Y\right)f\left(Y\right)f\left(\epsilon\right)\mathrm{d}X\mathrm{d}Y\mathrm{d}\epsilon\right\}
\thinspace\thinspace.
\end{equation}
As described earlier, we fix the joint distribution of the state space variables across cohorts, and so only the height distribution $\Gamma$ is cohort-specific in Equation \eqref{eq:targetcost}. We start $\Gamma_{1970}$ as mentioned using the actual height distribution in the year 1970 from Atole villages, and solve for subsequent reference point distributions following Equations \eqref{eq:mean_belief}, \eqref{eq:variance_belief}, and \eqref{eq:hmeasure}.

We first solve for $Z(\tau=0.1,\delta=0.9)$, when 10 percent of the poorest simulated households are provided with a 90 percent protein price discount. Then, for $\tau\in{(0.2,0.3,...,0.9,1.0)}$, we solve for the $\delta$ that minimizes the difference between $Z(0.1,0.9)$ and $Z(\tau,\delta)$:
\begin{equation}
\label{eq:deltaargmin}
\delta\left(\tau,Z\left(0.1,0.9\right)\right)=\arg\min_{\delta\in\left[0.01,...,1.00\right]}|Z\left(\tau,\delta\right)-Z\left(0.1,0.9\right)|
\end{equation}
Solving for the $\delta$ values following Equation \eqref{eq:deltaargmin}, we find that policies that provide 55, 50, 39, 30, 25, 21, 18, 16, 14 and 13 percent price discounts for 20, 30, 40, 50, 60, 70, 80, 90, and 100 percent of children, ranked from the poorest to the richest, cost approximately the same as $Z\left(0.1,0.9\right)$.\footnote{These budget-balancing price discounts differ minimally when estimates from different columns of Table \ref{tab:paramesti} are used. The specific numbers shown here are computed under $\sigma_{R_{y,v}}=3.5$. For consistency across comparisons, we use these price discounts for results when $\sigma_{R_{y,v}}=0.5$ as well.}

\subsubsection{Targeted Policies on Bottom Quintile ("Poorest") and Top Quintile ("Richest")}

We discuss in this section the impact of policies with different degrees of targeting/price-discounts on children from the bottom 20 "poorest" and top 20 percent "richest" of the income distribution. We compare targeting 20, 40, 60, 80 percent of the most impoverished children and universal policy. Price discounts at 55, 30, 21, 16, and 13 percent for each of the five policies are calculated to preserve budget balance. The different price discounts associated with different targeting policies are used for policy comparisons. The poorest children receive subsidies under all five policies, but the richest children only receive subsidies under the universal policy.

\begin{figure}[!ht]
	\centering
	\includegraphics[scale=0.24]{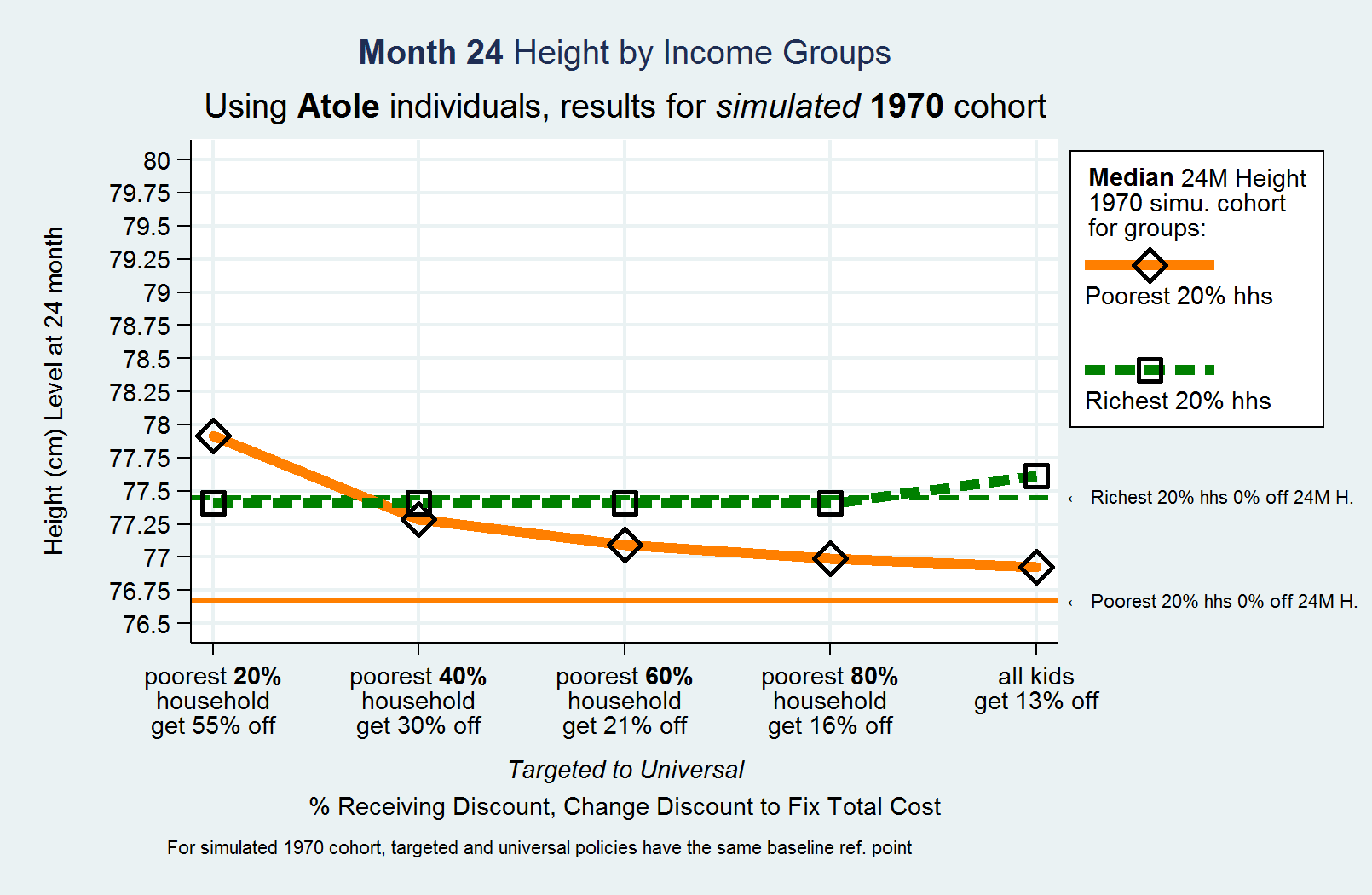}
	\caption{Effects of targeted policies on richest and poorest households (1970), $\sigma_{R_{y,v}}$ = 3.5~cm\label{fig:targety}}	
	\vspace{5mm}
	\centering
	\includegraphics[scale=0.24]{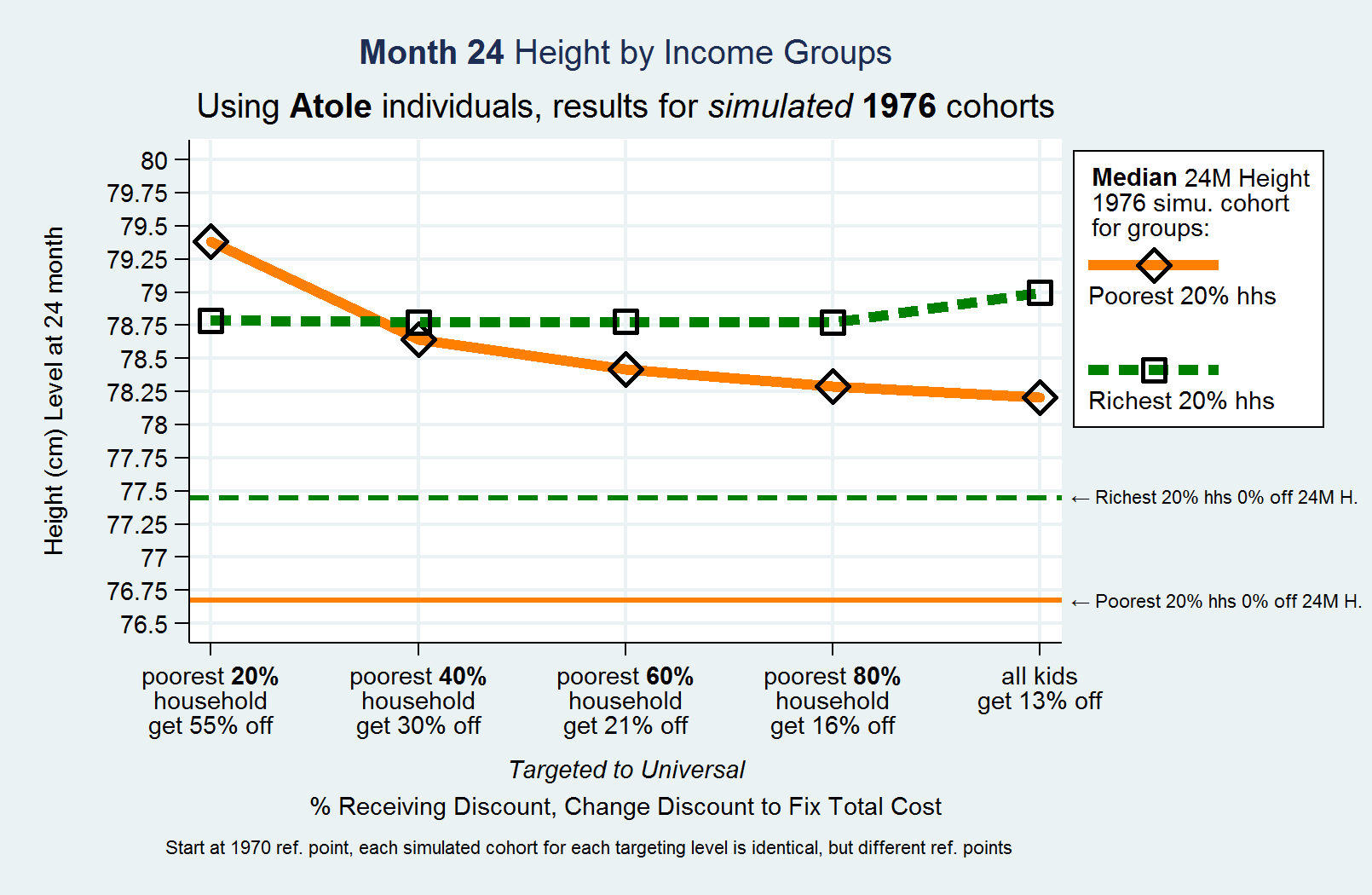}
	\caption{Effects of targeted policies on richest and poorest households (1976), $\sigma_{R_{y,v}}$ = 3.5 cm\label{fig:targety2}}
\end{figure}

\begin{figure}[!ht]
    \centering
	\begin{subfigure}[t]{1\textwidth}
        \centering
    	\includegraphics[scale=0.24]{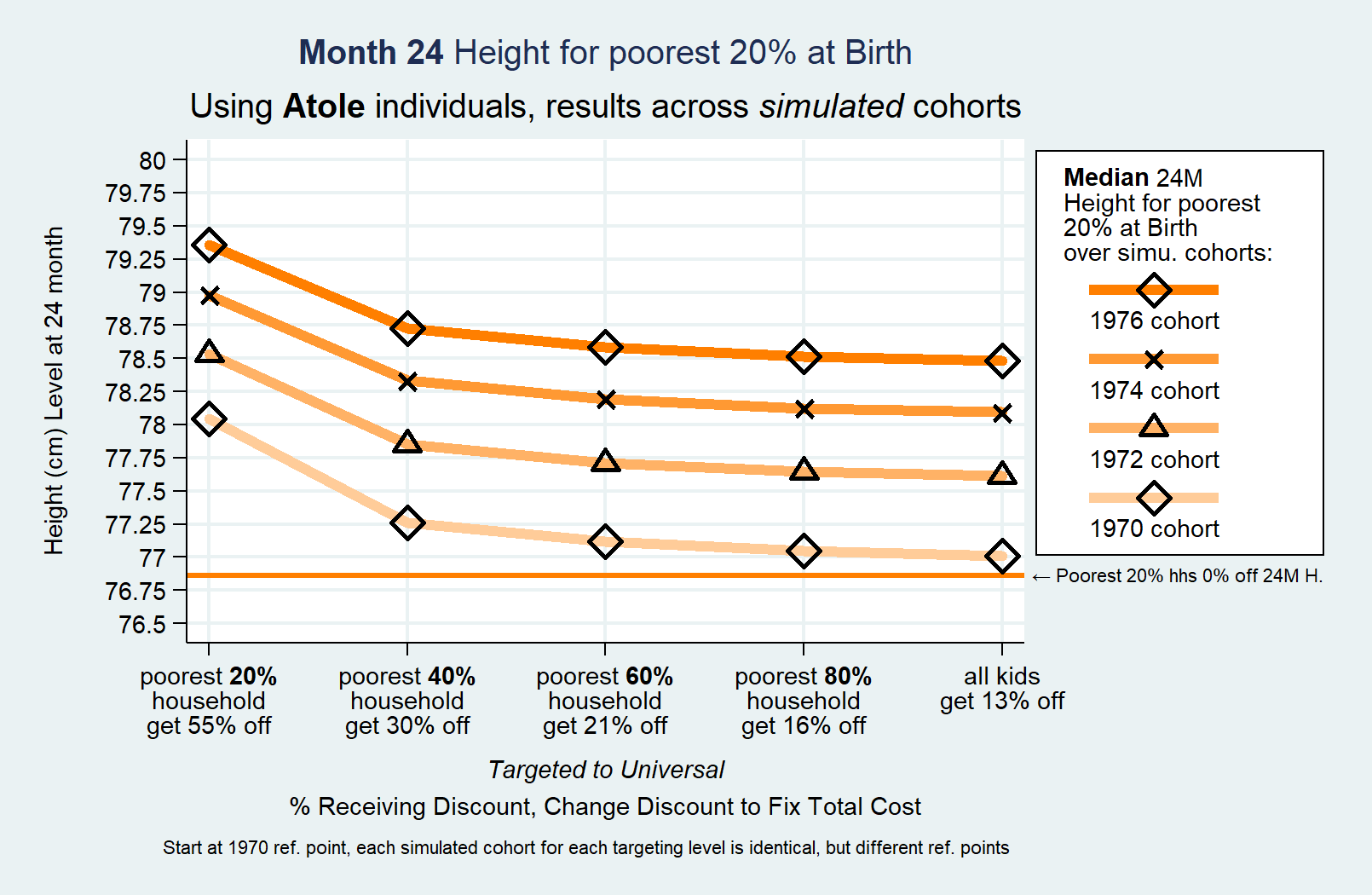}
    	\caption{$\sigma_{R_{y,v}}$ = 0.5 cm}
	\end{subfigure}
	\par\smallskip
	\begin{subfigure}[t]{1\textwidth}
	    \centering
    	\includegraphics[scale=0.24]{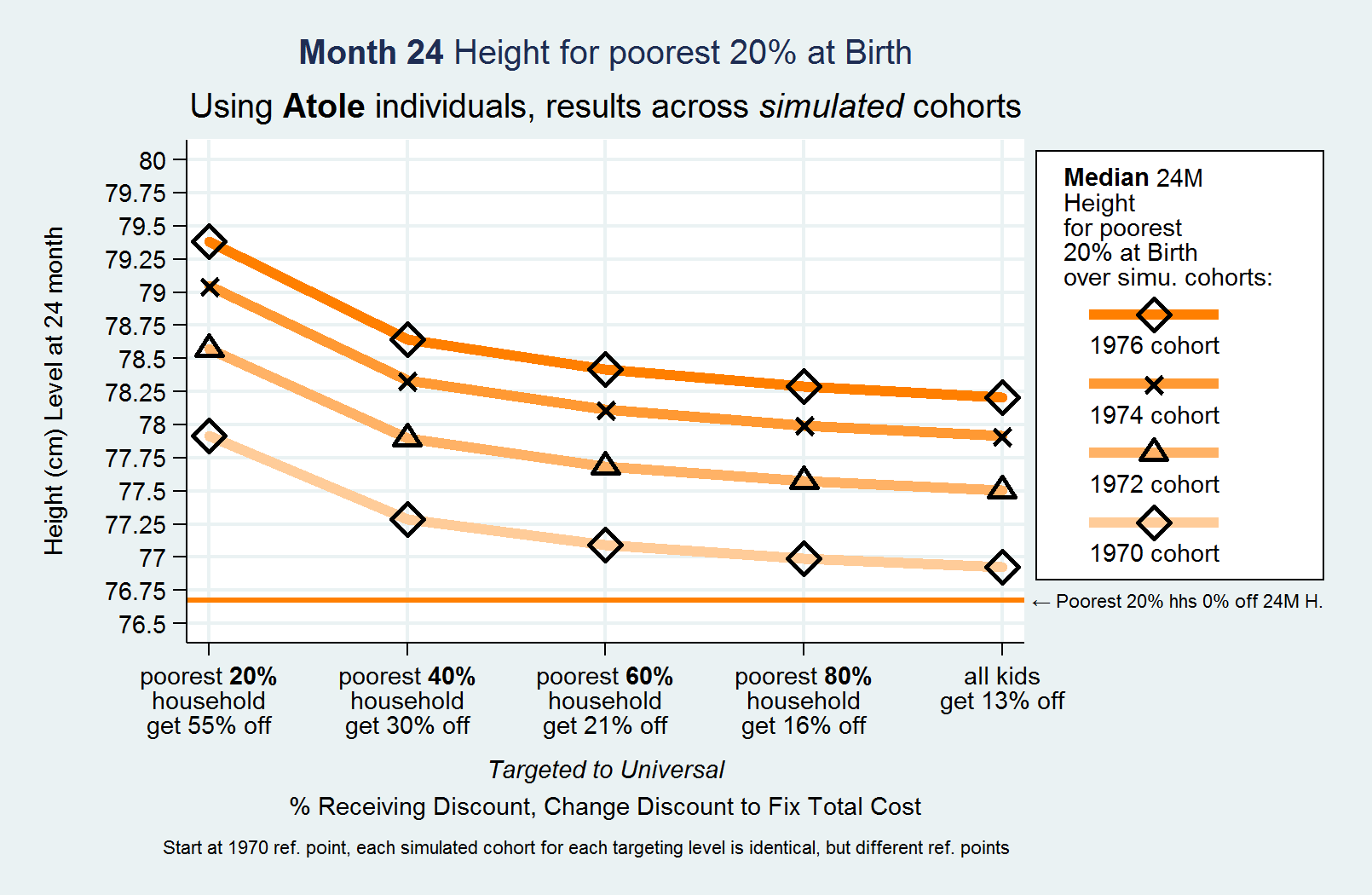}
    	\caption{$\sigma_{R_{y,v}}$ = 3.5 cm}
	\end{subfigure}
	\caption{Effects of targeted policies on poorest households}
	\label{fig:targetyt}
\end{figure}

We start the simulation with the 1970 reference points and the observed distributions. Figure \ref{fig:targety} presents results for the cohort born in 1970. Figure \ref{fig:targetyt} presents the differences in heights for the poorest children from the 1970, 1972, 1974, and 1976 cohorts. Figure \ref{fig:targety2} presents results for the 1976 cohort. Figure \ref{fig:targetyt} shows results using estimates from both $\sigma_{R_{y,v}}=0.5$ and $\sigma_{R_{y,v}}=3.5$. Due to space limitations, Figures \ref{fig:targety} and \ref{fig:targety2} are shown only for $\sigma_{R_{y,v}}=3.5$.
The orange solid (green dashed) line shows median heights for the poorest (richest) children in all figures. The flat horizontal lines show that in the absence of the price discount policy, median heights at month 24 for the poorest and the wealthiest children in the 1970 cohort were 76.7 and 77.4~cm, respectively (0.7~cm gap).\footnote{These correspond to median daily protein intakes of 19.0 and 22.1 grams for the poorest and wealthiest children.}


Figure \ref{fig:targety} shows that under targeted subsidies, the most impoverished children from the 1970 cohort experience a significant increase in heights, with median heights increasing by 1.3, 0.6, 0.4, and 0.3~cm when the most impoverished children receive 55, 30, 21, and 16 percent targeted discounts. With a 55 percent price discount, the most impoverished children's median height reaches 77.9~cm and is 0.5~cm higher than the median height for the wealthiest children. When both the richest and the most impoverished children receive 13 percent price discounts under the universal policy, median height increases more for the poorest (+0.25~cm) than the richest (+0.17~cm).


Figure \ref{fig:targetyt} shows similar results in panel a under $\sigma_{R_{y,v}}=0.5$ and in panel b under $\sigma_{R_{y,v}}=3.5$. Panel b of Figure \ref{fig:targetyt} shows that with 55 percent price discounts under the most-targeted policy, due to endogenous shifts in reference-point distributions, median heights increase significantly by 1.3, 1.9, 2.4, and 2.7~cm for the 1970, 1972, 1974, and 1976 cohorts of the poorest children, respectively.\footnote{Correspondingly, median protein intake increases by 23, 38, 50, and 58 percent for these cohorts.} On the other hand, the universal policy provides small height increases to the poorest initially, but effects amplify significantly across cohorts. Specifically, for the 1970 cohort of the poorest children, the most-targeted policy (55 percent discount) increases the median height by 1.3~cm, which is 4.8 times larger than the 0.26~cm median height increase induced by the universal policy (13 percent price discount). For the 1976 cohort of the poorest children, however, the effect of the most-targeted policy is only 1.8 times greater than the effect of the universal policy, each of which increases median heights by 2.7 and 1.5~cm, respectively.\footnote{Comparing the poorest children across cohorts, the increase in median height is 5.8 times larger for the 1976 cohort compared to the 1970 cohort (1.5 vs. 0.26~cm). The increase in median height for the richest children across cohorts is 2.2 times (2.7 vs. 1.25~cm). Additionally, for the 1970 cohort of the poorest children, 100 percent of the increases in height are due to first-period price-discount effects, but for the 1976 cohort of the poorest children, under 55, 30, 21, 15, and 13 percent price discounts, first-period price-discount effects only account for 46, 31, 24, 19, and 16 percent of the total effect of each policy.}

Figure \ref{fig:targety2} presents results for the 1976 cohort. Here we compare heights for the 1976 cohort to the heights at month 24 for the 1970 poorest and richest cohorts without subsidies. Across the five policies from most-targeted to universal, heights for the richest children increase by 1.3, 1.3, 1.3, 1.3, and 1.6~cm, equivalent to 48, 65, 76, 81, and 103 percent of the increases in heights for the poorest children, which are 2.7, 2.0, 1.7, 1.6, and 1.5~cm. The 1.3~cm median height increases for the richest children under the four targeted policies are due to the reference-point externality of the treatments on non-targeted individuals. The poorest children reach a higher median height (79.4~cm) than the richest non-targeted children (78.8~cm) only under the 55 percent price discount policy.

\subsubsection{Distributional Effects of Different Levels of Targeted Policies}

This section focuses on illustrating the potential distributional effects of targeting given reference points. A planner might be interested in maximizing some joint village welfare function that considers the entire distribution of height or moments of this distribution. We do not assume which percentiles or levels of height interest the planner. Given the specifics of our empirical setting, we present the full predicted distribution that contains information that might be of interest to a planner interested in early childhood height and health outcomes.

\begin{figure}[!t]
    \centering
	\begin{subfigure}[t]{1\textwidth}
        \centering
    	\includegraphics[scale=0.24]{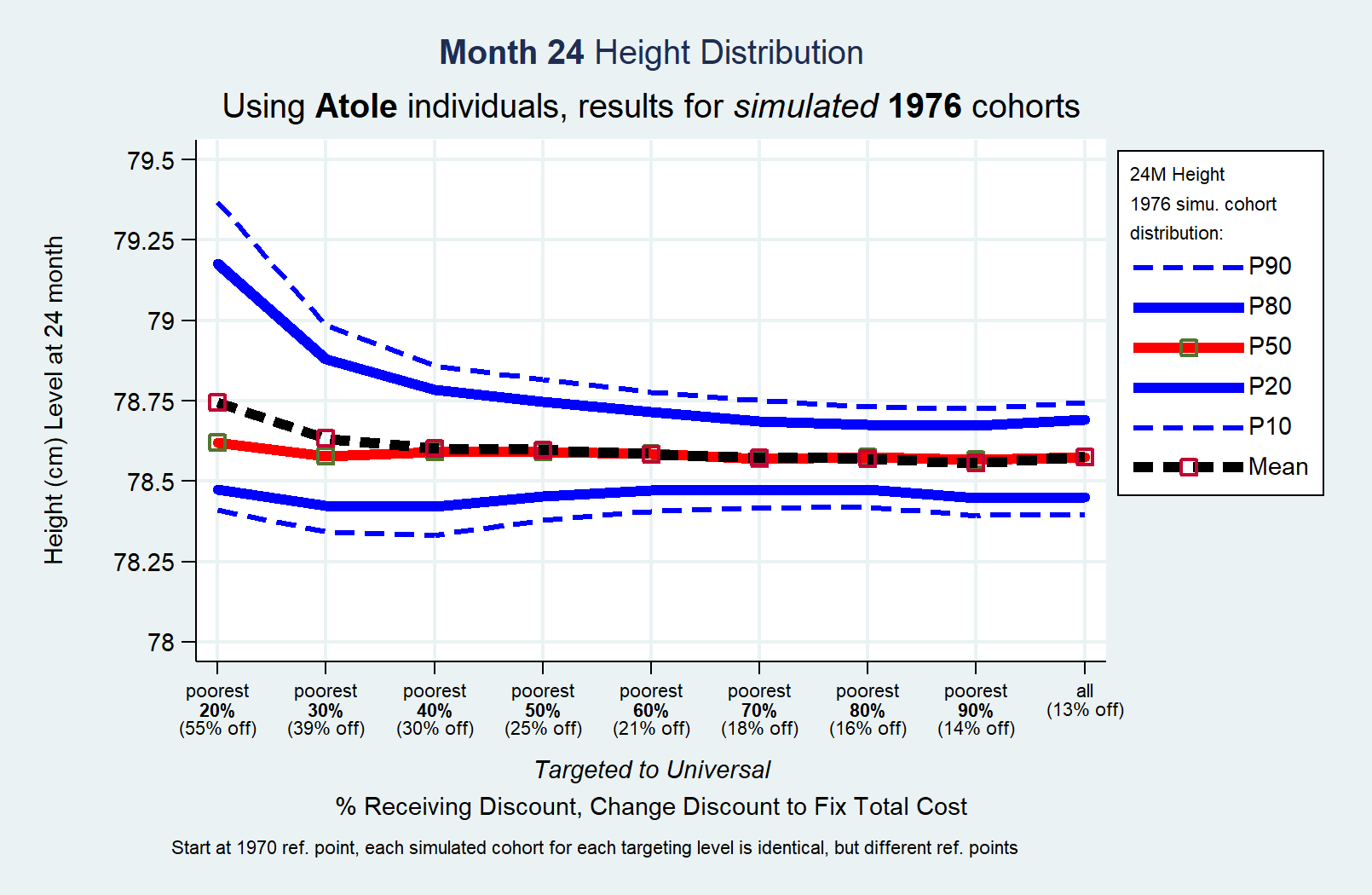}
    	\caption{$\sigma_{R_{y,v}}$ = 0.5 cm]}
	\end{subfigure}
	\par\smallskip
	\begin{subfigure}[t]{1\textwidth}
	    \centering
    	\includegraphics[scale=0.24]{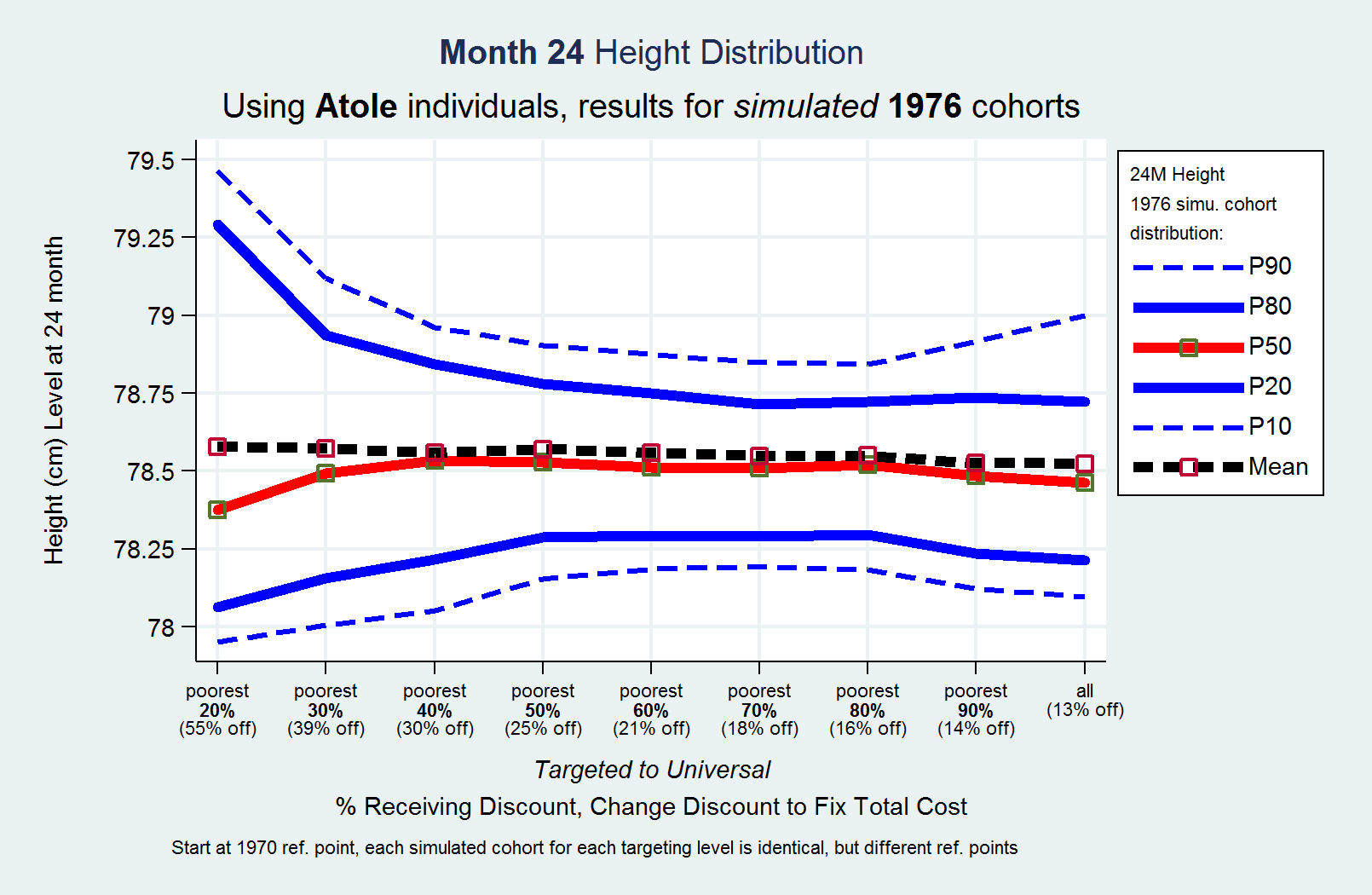}
    	\caption{$\sigma_{R_{y,v}}$ = 3.5 cm}
	\end{subfigure}
	\caption{Height Distribution from Targeted to Universal Subsidies}
	\label{fig:targetiqr}
\end{figure}

Figure \ref{fig:targetiqr} plots various percentile levels of overall -- including both targeted and non-targeted children -- heights of month-24 distributions\footnote{The distributional outcomes are presented without drawing from the measurement error distributions.} under each policy experiment. As shown in both panel a and panel b of Figure \ref{fig:targetiqr}, given our sample of individuals, mean heights are relatively constant across targeting levels, and variance is minimized when 70 percent of the poorest children are targeted to receive subsidies. Specifically, we analyze the effects of targeting from 20 to 90 percent of the poorest children at 10 percent intervals along with a universal policy. Price discounts at 55, 39, 30, 25, 21, 18, 16, 14, and 13 percent for each of the nine policies are calculated to be approximately budgetary constant.

The dashed black line in both panels of Figure \ref{fig:targetiqr} shows that the mean heights across policies differ by at most only 0.054~cm.\footnote{The highest mean in panel b, 78.58~cm, is achieved under targeting 20 percent of the poorest households, and the lowest mean, 78.52~cm, is achieved under the universal policy.} Our policy experiments shift a fixed amount of subsidies from one subset of individuals to another in the form of price discounts. Our findings suggest that these policies have linear effects on height.\footnote{One might suspect that the variations in means across policies would be large given the concavity of the production function: all else equal, one gram of protein transferred from those with high intake to those with lower intake should lead to a net gain in overall height. Here, however, given that households re-optimize with new subsidies, the increases in protein intakes are less than the amounts of protein transferred to the poor.} Who receives the transfers and how much they receive changes the relative heights among individuals, but not the overall average height significantly.

As policies shift price-discount intensity and recipients, there are significant variations in height distributions across policies. One measure of variation is the difference between the 10th and 90th percentiles of heights across policies, which is the difference between the dashed blue lines at the top and bottom of both panels of Figure \ref{fig:targetiqr}. In panel b, these gaps are 1.5, 1.1, 0.9, and 0.75~cm when 20, 30, 40, and 50 percent of the most impoverished children receive price discounts. The gaps are the tightest at 0.69, 0.66, and 0.66~cm under policies that provide 60, 70, and 80 percent of most impoverished children with price discounts;\footnote{In panel b, the s.d. of heights, at 0.40~cm, is also the smallest for the policy that targets 70 percent of most impoverished children.} the gaps widen again to 0.8 and 0.9~cm under the two most universal policies.

In our setting, the tightening of the height distribution's higher and lower percentiles drives the minimization of height variation when 70 percent of the most impoverished children are targeted. Under the most-targeted policies, children from the poorest households receive huge discounts and become the tallest children, driving the 90th percentile of height at the month-24 distribution up.\footnote{For the most-targeted policy that provides 55 percent price discounts to the children in the lowest quintile of income, these most impoverished children's heights increase significantly, and they move to the highest quintile of realized height in the month-24 distribution. As shown in Figure \ref{fig:targety}, the median heights of these most impoverished children under the 55 percent discount policy exceed the median heights of children from the wealthiest quintile of income. Hence this pushes the 80 and 90th percentile of the overall height distribution up under the most-targeted policies.} Concurrently, many children with below-median income do not receive subsidies under the most-targeted policies, and they push down the 10th and 20th percentiles of the height distribution. Second, under the universal policies, children from the wealthiest households receive price discounts, and they drive the 90th percentile height distribution up.\footnote{As shown in Figure \ref{fig:targety}, the smaller discount (13 percent discount under the universal policy) given to the wealthiest children allows them to achieve higher heights at month 24 than a more significant discount (16 to 21 percent under slightly more-targeted policies) given to children in the lower portions of the income distribution.} Concurrently, because subsidies for the poorest households are much lower than under more-targeted policies, they push the 10th and 20th percentiles of the height distribution lower. Consequently, we observe the broader distributions of heights under the most-universal and most-targeted policies, but a tighter distribution in the middle of Figure \ref{fig:targetiqr}.
\clearpage

\end{document}